\documentclass[pra,10pt,twocolumn,superscriptaddress]{revtex4}
\usepackage{amsmath}
\usepackage{latexsym}
\usepackage{amssymb}
\usepackage{graphics,epstopdf}
\usepackage{graphicx}
\usepackage[colorlinks=true, citecolor=blue, urlcolor=blue ]{hyperref}
\usepackage{float}
\usepackage{graphicx}
\usepackage{amsfonts}
\def\ket{\rangle}
\def\bra{\langle}

\begin{document}

\title{One-shot conclusive multiport quantum dense coding capacities}

\author{Chirag Srivastava}
\affiliation{Harish-Chandra Research Institute, HBNI, Chhatnag Road, Jhunsi, Allahabad 211 019, India}

\author{Anindita Bera}
\affiliation{Harish-Chandra Research Institute, HBNI, Chhatnag Road, Jhunsi, Allahabad 211 019, India}
\affiliation{Department of Applied Mathematics, University of Calcutta, 92 A.P.C. Road, Kolkata 700 009, India}
\affiliation{Racah Institute of Physics, The Hebrew University of Jerusalem, Jerusalem
91 904, Givat Ram, Israel}

\author{Aditi Sen(De)}
\affiliation{Harish-Chandra Research Institute, HBNI, Chhatnag Road, Jhunsi, Allahabad 211 019, India}

\author{Ujjwal Sen}
\affiliation{Harish-Chandra Research Institute, HBNI, Chhatnag Road, Jhunsi, Allahabad 211 019, India}

\begin{abstract}

We introduce a probabilistic version of the one-shot quantum dense coding protocol
in both two- and multiport scenarios, and refer to it as conclusive quantum 
dense coding. Specifically, we analyze the corresponding capacities of shared states between two, three, and more qubits, and two qutrits. We identify 
cases where Pauli and generalized Pauli operators are not sufficient as encoders to attain the optimal one-shot conclusive quantum dense coding capacities. We find that there is a rich connection between the capacities, and the bipartite and multipartite entanglements of the shared state.


\end{abstract}

\maketitle
\section{Introduction}
\label{Intro}


Any communication protocol typically involves two parties, a sender and a receiver. A sender 
%
encodes the information in a physical system, sends it through a physical channel, and then the other party -- the receiver -- performs a decoding. In this paper,
we are interested in a class of communication schemes that deal with transfer of classical information through quantum states shared between distant parties, broadly known as quantum dense coding (DC). The first instance of a DC protocol was provided by Bennett and Wiesner~\cite{ben} in 1992,
in which they proposed to use a single copy of a certain shared state between two parties, say, Alice (sender) and Bob (receiver), and a noiseless quantum channel between them, as resources. Utilizing these resources, Alice is able to transfer an amount of classical information that is higher than the corresponding ``classical limit"~\cite{eijey}, for certain shared states.
It turns out that such an advantage over classical protocols is obtained due to the quantumness present in a shared quantum state, named as entanglement 
\cite{4h,Das-Chanda}. Over the years, 
it has also been realized that entangled states are not only useful for sending classical information but are also beneficial for different quantum information processing tasks \cite{4h}, and such tasks have been performed with different physical systems like
%
%
 photons~\cite{ph1, ph2, ph3, ph4, ph5, ph6, ph7} and ions~\cite{ion1, ion2}  (see also 
 \cite{sup-qubit, 
 optical-lattice,
 NMR, NMR1, NMR2}).
%
In 1996, Mattle \emph{et al.} performed an experiment to demonstrate the quantum DC protocol by using the polarization degree of freedom of 
photons~\cite{mattle-exp}. See also~\cite{dc-exp2,dc-exp3,dc-exp4,dc-exp5}.

There is a significant body of work on generalizations of the original DC protocol
in which the unitary encoding at the sender's node is on a single copy of the state while the decoding process of the receiver involves multiple copies, and refer to them as the generalized DC protocols. See, e.g.,~\cite{dense1,dense2,dense3,dense4,dense5,ziman,barenco,laden,hiro,piani,holevo, schum,sir,Braunstein-DC,Zhang-DC,hao,hao-multiparty,roger}. See also \cite{swilde}.
In a different direction,
Mozes \emph{et al.}~\cite{jon} have introduced a DC protocol, called 
``deterministic dense coding" (DDC) protocol, where both encoding and decoding are performed at the single-copy level of the shared quantum state between Alice and Bob, and Alice is able to send classical data to Bob in an error-free mode, by using non-maximally entangled pure states. Several interesting questions including generalization of a DDC scheme to a multiparty scenario have been addressed in recent years
\cite{ddc-ref,ji,sap}.

In this paper, we introduce 
the concept of $conclusive$ dense coding (CDC), in which Alice uses a single copy of a pre-shared non-maximally entangled pure (quantum) state with Bob to send $complete$ classical messages unambiguously. The protocol, which, like the DDC scheme, is a single-shot strategy,  allows for some error (because of its probabilistic nature), but the users know when an error has been committed, and for a fixed shared state, 
the aim 
is to devise a quantum strategy (encoding and decoding) to minimize the error. And in case an error is not committed, Alice is successful in sending the $complete$ classical message what she intended to send.

To clarify the meaning of sending a $complete$ classical message, suppose that Alice intended to send information about the color and the softness of the ball. And if a protocol, sometimes, allows to send information about its color only, then it is not $complete$ information. In the CDC protocol, we equate receipt of such $partial$ information with receipt of \emph{no} information. The situation is similar to e.g. certain online transactions where \emph{both} card verification value (``CVV") and one-time password (``OTP") are needed, and possessing any one is useless for the task. 
Previous works on probabilistic versions of the dense coding protocol can be found in~\cite{barenco,ddc-ref,hao,hao-multiparty,roger}, although none of the 
schemes include the 
stipulation
of \emph{complete} information transfer.

 An important feature of utilizing CDC instead of the related DDC protocol for a shared state of two $d$-dimensional quantum systems is that the latter is not useful for sending a (classical) message that has more than $d^2-2$ values if the shared state is not maximally entangled~\cite{ji}, while the former is. Probabilistic versions of the DC protocols and their capacities have often been calculated in previous works by using generalized Pauli operators as encoders. Such encoding may not provide the optimal capacity, as is the case for several of the scenarios considered here. 
 We also comment on the relation of the CDC capacity with the entanglement in the shared state.

 We subsequently move over to the case of multiport conclusive quantum dense coding
with 
two senders and a single receiver sharing a genuine pure tripartite entangled state. Recently, it was shown that generalized Greenberger-Horne-Zeilinger (GHZ)~\cite{GHZ} states are not useful for DDC except at the GHZ point \cite{sap}. Here we reveal that quantum advantage of CDC for every pure state, entangled in the senders : receiver bipartition. We also report that there is a marked difference between the generalized GHZ and the generalized W~\cite{dur,ZHG,mam} states, in terms of the relation between their multiparty conclusive quantum dense coding capacities and (bipartite and multipartite) entanglement contents. 

We arrange the paper in the following way. In Sec.~\ref{motivation}, we begin by recounting the original DC protocol and its generalization to non-maximally entangled states. We then describe the one-shot conclusive DC scheme, including the motivation to consider the same.
Next, in Sec.~\ref{settings}, we frame the capacity of the CDC protocol 
for arbitrary shared pure states.
We illustrate the CDC capacities for arbitrary two-qubit and two-qutrit pure states 
in Sec. \ref{bipartitePDC}. In Sec.~\ref{multipartyPDC}, we extend our protocol to a multiple senders and single receiver picture, who are situated at different locations, followed by concluding remarks in Sec.~\ref{conclusion}.


\section{Conclusive quantum dense coding}
\label{motivation}

We begin by providing a brief description of the 
quantum dense coding protocol, discovered by Bennett and Wiesner~\cite{ben}. Two parties, named as Alice and Bob, share a two-qubit system in the singlet state, $|\psi^-\ket=(|01\ket-|10\ket)/\sqrt{2}$, with $|0\ket$ and $|1\ket$ being orthogonal eigenvectors of the Pauli-$\sigma_z$ operator. Alice applies the set of local unitaries $\{\mathbb{I},\sigma_x,\sigma_y,\sigma_z\}$, where $\mathbb{I}$ is the identity operator on the qubit space and $\sigma_i$, $i=x, y, z$, are the Pauli spin matrices, with equal probabilities, on her part of $|\psi^{-}\rangle$. She then sends her part to Bob using a quantum channel that can noiselessly carry a qubit, thereby creating an ensemble of the states  $\{|\psi^{\pm}\rangle,|\phi^{\pm}\rangle \}$ 
\cite{bell-eita}, with equal probabilities, at Bob's laboratory.
As these states are orthogonal, Bob can, in principle, distinguish between them with certainty, and hence
Alice in this case can send $\log_2 4 =2$ bits of classical information, or equivalently, a 4-valued classical message, to Bob. When the shared state is not entangled, Alice can send only 1 bit, i.e., a two-valued classical message, to Bob, and is the ``classical limit" in this case. 

Instead of possessing a singlet, or a state local unitarily connected to it, which provides the highest capacity for dense coding, 
a non-maximally entangled state can be shared between Alice and Bob, and may turn out to be the natural vehicle for transferring classical data at a certain stage of a quantum machine. Suppose, therefore, that Alice and Bob share an arbitrary quantum state $\rho_{AB}$ on $\mathbb{C}^{d}\otimes\mathbb{C}^{d}$, and  Alice wants to send the classical information  $i$ to Bob. Depending on the classical message $i$, Alice performs the unitary operation $U_i$ on her part of the shared state. We now suppose that Alice and Bob have at their disposal, a quantum channel that can noiselessly transfer a $d$-dimensional quantum state. Using this channel, Alice sends her part of $\rho_{AB}$, after the action of $U_i$, to Bob. Bob therefore possesses a quantum ensemble, $\{p_i, \rho_{AB}^i\}$, and his task is to find as much information as possible by performing a quantum measurement on the ensemble. The information gathered after performing a measurement by Bob can be quantified by the mutual information~\cite{cov} between the measurement results and the index $i$. This mutual information is dependent upon the measurement strategy of Bob and the encoding parameters of Alice, and a maximization over these strategies and parameters will provide us the amount of classical information that can be sent from Alice to Bob. This maximum is bounded above by ~\cite{laden,barenco,ziman,hiro,piani,holevo,schum,sir}
\begin{equation}
\label{Eq.2}
C_a(\rho_{AB})=\log_2{d}+\mbox{max}\{S(\rho_B)-S(\rho_{AB}),0\}
\end{equation}
bits, where $S(\sigma)=-\text{tr}(\sigma \log_2 \sigma)$ is the von Neumann entropy~\cite{wehrl} of $\sigma$ and $\rho_B=\text{tr}_A(\rho_{AB})$ is the reduced density matrix of Bob's subsystem. This bound can be asymptotically attained, assuming an encoding that is product over different uses of the quantum channel~\cite{laden,holevo,schum}. $C_a$ is usually referred to as the DC capacity of $\rho_{AB}$. It is measured in bits, being a consequence of the choice of 2 as the base of the logarithms appearing in Eq.~(\ref{Eq.2}). We will continue to use the same base of logarithms appearing in this paper, so that the other capacities and mutual information are also measured in bits, even though the unit will often be kept silent in the definitions. It is important to note here that the first term in Eq.~(\ref{Eq.2}), $\log_2{d}$ bits (i.e., a $d$-valued classical message), provides the ``classical limit", being defined as the maximum number of bits that can be transferred through a quantum channel that noiselessly transfers a $d$-level quantum system, without the additional resource of any pre-shared entangled state between the laboratories at the ends of the channel. The term $\mbox{max}\{S(\rho_B)-S(\rho_{AB}),0\}$ is the ``quantum advantage" in DC with the state $\rho_{AB}$, being non-zero for states with $S(\rho_B)>S(\rho_{AB})$, and being equal to entropy of entanglement~\cite{entropy1,entropy2} for pure shared states. The quantum DC protocol has been generalized in several directions. In particular, Braunstein \emph{et al.}~\cite{Braunstein-DC} and Zhang \emph{et al.}~\cite{Zhang-DC} have proposed DC protocols for continuous variables and the authors in Refs.~\cite{hao-multiparty,dense5,sir} have generalized the DC protocol to the multipartite case.

 At this point, it is important to stress that the DC capacity, $C_a$, given in Eq.~(\ref{Eq.2}), is the maximization of the mutual information between the message sent (the index $i$ of the ensemble $\{p_i,\rho_i\}$) and the message received (the outcome of measurement by Bob). While this is an important quantity in theory and practice of information transmission, its value does not ensure that the retrieved 
 information is the same as that sent, unless the mutual information touches  $2\log_2{d}$ bits. If the maximal mutual information touches the $2\log_2{d}$ bits limit, it is possible to send a $d^2$-valued classical message without any error. However, if the maximal mutual information is lower than 
  \(2\log_2 d\), one cannot ensure an error-free transmission of a classical message that has more than \(d\) values (but less than \(d^2\)). A $d$-level classical message can always be sent exactly, as discussed before, since we have a noiseless qudit channel at our disposal. Also the DC capacity, $C_a$, is achieved, in general, only when an infinite number of copies of the initial state, required for decoding, is shared.
 
It is therefore interesting to consider another approach of DC which deals with \emph{conclusive} transfer of classical information through a single copy of a pure shared quantum state. 
 The protocol is \emph{conclusive} in the sense that when an $N$-valued classical message, encoded in an $N$-element ensemble on $\mathbb{C}^d\otimes\mathbb{C}^d$, is obtained by Bob, with $N>d$, it is possible to unambiguously distinguish between the $N$ options with a non-zero probability. Note that we must have $N\leqslant d^2$. The distinguishing protocol is ``unambiguous'' in the sense that while there may be errors in discrimination between the $N$ elements of the ensemble, we are able to know when an error has been committed. In case the discrimination succeeds with unit probability, the protocol is termed as 
 ``deterministic dense coding" ~\cite{jon,ddc-ref,ji,sap}. It is clear therefore that deterministic dense coding is a special, and important, instance of conclusive dense coding.

In a different approach to probabilistically send classical information over quantum channels, as considered by Hao \emph{et al.}~\cite{hao} (see also \cite{barenco,hao-multiparty}), 
\emph{partial} transfer of information was allowed from Alice to Bob.
To describe it, let us consider an example where the information about whether a ball 
is either hard or soft, and is either black or blue, is to be sent from Alice to Bob, implying that there are two bits of information that are to be sent. Suppose now that the sender is able (or decides) to communicate the information \emph{only} about the color of the ball. In the scenario considered by Hao \emph{et al.}, this constitutes sending one bit of information. This is different from the approach in conclusive quantum dense coding, as followed in this paper, since we consider this incidence of sending partial information (when only the color of the ball is sent), as equivalent to \emph{not} sending any information. There exist situations in which, until and unless one has the full information about the system, it is effectively a ``failure". For example, consider the situation where a thief has stolen a person's debit card and wants to do an online transaction with a merchant. But certain cards have an additional level of security, e.g. an one-time password sent to a pre-decided mobile phone, in absence of which, the stolen card is, happily, useless for any transaction.


\section{Capacity of conclusive quantum dense coding}
\label{settings}
  
  Let us describe the conclusive dense coding protocol in a little more detail. Suppose that two parties, Alice and Bob, share the state $|\Psi\ket$ of $\mathbb{C}^d\otimes\mathbb{C}^d$, and also
  possess a quantum channel that can noiselessly transfer a $d$-dimensional quantum system. With this doublet of resources, the task is to transfer an $N$-level classical message unambiguously, for which they adopt the following strategy. Alice applies an $arbitrary$ ensemble of unitaries, $U_i$, defined on $\mathbb{C}^d$, with respective probabilities, $p_i$, creating the ensemble $\{p_i,|\Psi_i\ket\}$, where $|\Psi_i\ket=U_i\otimes\mathbb{I}|\Psi\ket$, $i=1,2,\ldots,N$. Here, $\mathbb{I}$ denotes the identity operator on $\mathbb{C}^d$. Without loss of generality, we can choose $U_1=\mathbb{I}$, as this fixes the local basis on Alice's side. Alice subsequently uses the noiseless quantum channel to send her part of the shared state to Bob, so that the two-system ensemble $\{p_i,|\Psi_i\ket\}$ is entirely in Bob's laboratory.   
  
  If  $|\Psi_i\ket$ are mutually orthogonal, they can, in principle, be distinguished exactly by Bob, and the entire $N$-level classical message will be obtained by Bob, with unit probability. To attain orthogonality, we need to find the corresponding unitaries that lead to this orthogonality, and this is the program of deterministic dense coding. If, however, the $|\Psi_i\ket$ are not mutually orthogonal, they can still be unambiguously distinguished with non-unit probability, which is non-zero if the $|\Psi_i\ket$ are linearly independent~\cite{Duan}. More precisely, Ref.~\cite{Duan} shows that the set of linearly independent states $|\Psi_1\ket, |\Psi_2\ket, \ldots, |\Psi_N\ket$ can be recognized with probabilities   $\gamma_1, \gamma_2, \ldots, \gamma_N$, respectively, if and only if
\begin{equation}
\label{cond1}
X^{(1)}-\Gamma \geqslant 0,
\end{equation} 
 where $X^{(1)}=\bra\Psi_i|\Psi_j\ket$ and $\Gamma=\gamma_i\delta_{ij}$. Due to the fact that the states $|\Psi_i\ket$ appear, in the ensemble in possession of Bob after Alice's use of the noiseless quantum channel, with probabilities $p_i$, the total probability of successfully recognizing the states is given by $\sum_{i=1}^{N}p_i\gamma_i.$ 


In order to obtain the capacity of conclusive dense coding, let us first define the classical mutual information, $\tilde{I}(i:m),$ between the message, $i$, which can take values, say, $1,2,\ldots,N$, sent with probabilities $p_i$, and the outcome \(m\) of a measurement,
having 
values, say, $1',2',\ldots,M'$, received with the probabilities $q_m$, as~\cite{cov}
\begin{equation}
\label{Eq.4}
\tilde{I}(i:m) = H(\{p_i\}) - \sum_{m=1'}^{M'} q_mH(\{p_{i|m}\}_i),
\end{equation} 
where $p_{i|m}$ is the conditional probability of the classical message $i$ for the outcome 
$m$, and 
$H(\{\zeta_j\})=-\sum_{j} \zeta_j\log_2 \zeta_j$ is the Shannon entropy of the probability distribution ${\zeta_j}$.

However, this standard definition and conceptualization of classical mutual information between two classical random variables is applicable to the case at hand, i.e. in the setting of conclusive dense coding, only after certain crucial modifications. First of all, we will only consider those strategies as valid conclusive dense coding protocols, for which every measurement outcome $m$ unambiguously corresponds to only one classical message $i$ or does not provide any information. And furthermore, we only consider strategies for which every $i$ corresponds to at least one $m$ that occurs with a non-zero probability in the measurement. Consider, for example, a strategy in which we have $i=$1, 2, 3 and $m=$1, 2, 3, such that $m=1$ implies either $i=1$ or $i=2$, $m=2$ implies $i=3$, and $m=3$ implies that nothing can be deduced about $i$. This strategy is discarded because the outcome $m=1$ does not unambiguously imply a particular $i$. Consider another strategy where we have $i=1$, 2 and $m=1$, 2, 3 such that $m=1$ implies $i=1$, $m=2$ implies $i=2$, and $m=3$ implies that nothing can be deduced about $i$, and moreover, $q_2=0$. This strategy is also disregarded because $i=2$ is never implied by any outcome. So, for an arbitrary but fixed $m$, the conditional probability $p_{i|m}$ has the following properties: 
\begin{itemize}
\item Either $p_{i|m}=1$ for a particular $i$, and $=0$ for the rest. Or $p_{i|m}=0$ for all $i$.
\item For every $i$, there exists at least one $m=m'$, with $q_m\neq 0$, such that $p_{i|m'}=1.$
\end{itemize}
Therefore, the outcomes can be divided into two categories - ones for which there is no remaining disorder (uncertainty) about the ensemble, and ones for which there is no or incomplete depletion of disorder about the same. Let us denote these sets by $\mathcal{S}$ and $\mathcal{F}$ respectively. Let the corresponding mutual information be denoted by $I(i:m)$. For cases when $m \in \mathcal{F}$, we assume that no information has been transmitted, so that the remnant disorder is the full disorder, viz. $H(\{p_i\})$. Therefore,   
\begin{eqnarray}
I(i:m)&=&H(\{p_i\})-\sum_{m\in \mathcal{S}} q_m \times 0 - \sum_{m\in \mathcal{F}} q_m H(\{p_i\})	\nonumber	\\
&=&(1-\sum_{m\in \mathcal{F}} q_m)H(\{p_i\}).	\nonumber
\end{eqnarray}
Now, $1-\sum_{m\in \mathcal{F}} q_m$ is exactly equal to $\sum_{j=1}^N  p_j\gamma_j,$ that has already appeared earlier. Therefore, we get
\begin{equation}
\label{Eq.6}
I(i:m)= \sum_{j=1}^N  p_j\gamma_j H(\{p_i\}).
\end{equation}   

 The free parameters of the function $I(i:m)$, for a given shared state and a given quantum channel, are the $p_i$ and $U_i$, with $\gamma_i$ being functions of $U_i$. The conclusive dense coding capacity, $C_N$,  for sending an $N$-level classical message, is defined as the maximum of $I(i:m)$ over the space of free parameters. We further assume that the classical messages appear with equal probabilities, i.e., $p_i=\frac{1}{N}$. Hence, the conclusive dense coding capacity for sending an equiprobable $N$-valued classical message, given a shared pure quantum state in $\mathbb{C}^d\otimes\mathbb{C}^d$ and a noiseless $d$-dimensional quantum channel, is given by
 \begin{equation}
 \label{Eq.7}
 C_N (|\Psi\rangle)= \max_{U_i}\sum_{i=1}^N \frac{\gamma_i}{N} \log_2{N},
 \end{equation}
 where the $\gamma_i$ are constrained by the relations in (\ref{cond1}).

 For an $N$-level classical message with $N>d$, one cannot send the message conclusively, using an unentangled shared pure state. This is because, any set of $N$ states that the sender can form (using local unitaries), will become linearly dependent, if the parties share an unentangled pure state. So the ``classical limit" is zero for $N$-level classical messages with $N>d$, and we say that there is a ``quantum advantage" when $C_N$ is strictly higher than zero. We define the CDC 
 capacity for a given state \(|\Psi\rangle\) as 
\begin{equation}
\label{Eq.8}
C(|\Psi\rangle)=\max_{N} C_N,
\end{equation} 
where the maximization is performed over all \(N>d\). For \(N\leq d\), one 
can ignore the shared state and reach a capacity of \(\log_2N\)
for all \(|\Psi\rangle\).

 There are several works on quantum dense coding, e.g. in  Refs.~\cite{barenco,jon,roger}, in which a ``standard encoding scheme'' utilizing generalized Pauli operators have been used. Generalized Pauli operators are given by $U_{mn}=X^m Z^n$, where $X$ and $Z$ are defined by their actions on the elements $|l\ket$ of an orthonormal basis of the Hilbert space $\mathbb{C}^d$:
\begin{eqnarray}
\label{Eq.7}
X|l\ket &=& |l\oplus 1\ket, \nonumber\\
Z|l\ket &=& e^{{2 \pi i l}/d}|l\ket.
\end{eqnarray}
Here $\oplus$ denotes addition modulo $d$, and $m,n=1, 2,\ldots, d$. 
The conclusive dense coding capacity obtained 
by
using the standard encoding scheme, 
for sending an $N$-valued classical message, is denoted as \(C_N^P\), and the maximum of these capacities over $N$ as $C^P$. 

\begin{center}
\begin{figure}[tb]
\includegraphics[width = 0.4\textwidth]{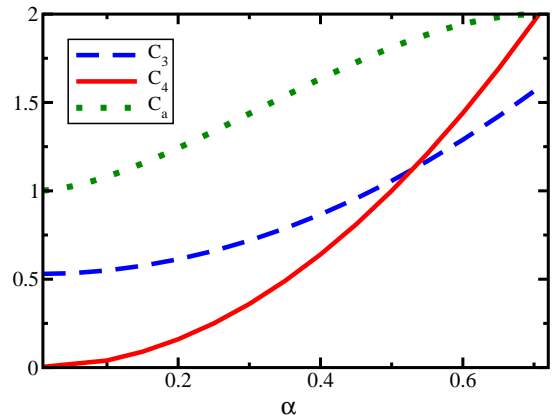} 
\caption{ Conclusive quantum dense coding capacity for two-qubit pure states. The CDC capacity $C_N$ for $N=3$ and 4, and 
the capacity $C_a$, in bits, are plotted with respect to the state parameter $\alpha$ in the two-qubit scenario. 
  The  capacity $C_a$ is an upper bound to the CDC capacities. All entangled states are conclusively dense codable, i.e., $C_N>0$  whenever $\alpha \neq 0$. The capacity, $C$, maximized over $N=3, 4$, is $C_3$ when $\alpha$ is small, but as $\alpha$ increases, $C$ becomes $C_4$ after $\alpha \approx 0.53$. The horizontal axis is dimensionless.
  }  
\label{fig.1}
\end{figure}
\end{center}

\section{Properties of Bipartite \(\mbox{CDC}\) capacity} 
\label{bipartitePDC}   
In this section,  we investigate the behavior of 
the CDC capacities for arbitrary pure states in $\mathbb{C}^2 \otimes \mathbb{C}^2$, and then 
in 
$\mathbb{C}^3 \otimes \mathbb{C}^3$, with respect to state parameters.

\subsection{CDC protocol in $\mathbb{C}^2 \otimes \mathbb{C}^2$}
\label{2-2 system}

Let us consider that an arbitrary pure state $|\Psi^{(2)}\ket=\alpha|00\ket+\sqrt{1-\alpha^2}|11\ket$ in $\mathbb{C}^{2}\otimes \mathbb{C}^{2}$, shared between Alice and Bob.  Without loss of generality, we assume that $\alpha$ is real and that $0\leqslant \alpha \leqslant \frac{1}{\sqrt{2}}$. The absence of a relative phase in $|\Psi^{(2)}\ket$ does not affect the generality of the considerations. We now discuss about the conclusive dense coding capacity of this state.

 In Fig.~\ref{fig.1}, we depict the capacity $C_N$, for $N = 3$ and $4$, with respect to the state parameter $\alpha$. Clearly, all such states, whenever entangled, provide an advantage over the case when there is no entanglement for being used as a resource in conclusive quantum dense coding, as their CDC capacities lie beyond the classical limit. The constrained optimizations here, and further on in the paper, are performed by non-linear optimization procedures \cite{nlopt}.

 For the maximally entangled states, i.e., for \(\alpha=1/\sqrt{2}\),  $C_N = \log_2 N$, because all of the efficiencies of identification ($\gamma_i$) become unity. Clearly, in this scenario, DDC can be executed. But for non-maximally pure entangled states, this situation does not arise and hence DDC is not possible in this case for both $N=3$ and 4.  So, we are able to reproduce the result of Mozes \emph{et al.}~\cite{jon} in which they showed that the entire family of pure states in $\mathbb{C}^2 \otimes \mathbb{C}^2$ except the maximally entangled state is not useful for DDC. However, the CDC capacity remains higher than the classical limit, whenever the states are entangled.    
 
We also compute the CDC capacity, $C_N^P$, using Pauli operators. For $N=3$ and 4, they are given by the following analytical form (see Appendix):
\begin{equation}
\label{Eq.9}
C_N^P=\frac{4-N+2\alpha^2(2N-4)}{N}\log_2{N}.
\end{equation}
An interesting fact is that the CDC capacities obtained using non-linear optimization over $arbitrary$ unitaries matches with the ones obtained using Pauli operators:
\begin{equation}
\label{Eq.10}
C_N=C_N^P.
\end{equation}

Note that $C_4$ is continuous for $\alpha \in$ $[0,\frac{1}{\sqrt{2}}]$, and $C_3$ is continuous for $\alpha \in$ $(0,\frac{1}{\sqrt{2}}]$. However, $\lim_{\alpha \to 0^+}C_3(\alpha)$ is not the same as $C_3(\alpha=0)$. $C_3(\alpha=0)$ is clearly vanishing, as for $\alpha=0$, the shared state is the product state $|11\ket$, which, when rotated by Alice can produce at most two linearly independent states that can never be used to conclusively transfer a 3-valued classical message. But, if $\alpha \neq 0$, 
although
close to zero, $|\Psi^{(2)}\ket=\alpha|00\ket+\sqrt{1-\alpha^2}|11\ket$ can be transformed by Alice to the ensemble consisting of $|\Psi^{(2)}\ket$, $\sigma_x \otimes \mathbb{I}|\Psi^{(2)}\ket$, and $\sigma_y \otimes \mathbb{I}|\Psi^{(2)}\ket$, 
which can be conclusively recognized with probabilities $\gamma_1=1, \gamma_2=\gamma_3=2\alpha^2$, respectively. This leads to a conclusive transfer of  $\frac{1+4\alpha^2}{3}\log_2{3}$ bits, which is actually the optimal one, as shown in 
Eqs.~(\ref{Eq.9}) $\&$ 
(\ref{Eq.10}). And for $\alpha \to 0^+$, this optimal transfer of information tends to a non-zero value.

Let us now consider the CDC capacity maximized over $N>2$. From Fig.~\ref{fig.1}, it is clear that the CDC capacity ($C$)
is $C_3$ for a certain range, viz. $0<\alpha \lesssim 0.53$, and $C_4$ for the rest,
viz.
$0.53\lesssim \alpha \leqslant \frac{1}{\sqrt{2}}$. 
Note that \(\alpha \approx 0.53\) (accurate up to the second decimal point) 
is the point where \(C_3\) and \(C_4\) intersect.

 The 
 capacity, $C_a=1 + S(\rho_B)$, is also plotted with respect to $\alpha$ in Fig.~\ref{fig.1}. Clearly, $C_a$ is an upper bound of all single-shot capacities, viz. $C_3\) and \(C_4$. Note that $C_a$ differs from entanglement entropy (i.e., the von Neumann entropy of the local density matrix of the shared pure state) just by an additive scalar in the case of pure bipartite shared states that we are considering, and is a strictly increasing function of the state parameter $\alpha$. 
 Interestingly, we observe that $C_3$ and $C_4$, too, are strictly increasing functions of the state parameter, thereby implying that they are 
 strictly increasing functions of the  entanglement of the shared state.
 While entanglement entropy is the asymptotic entanglement measure for pure two-party quantum states~\cite{benpop}, $\alpha^2$ and $\alpha$ are also entanglement measures (or monotones) for single-copy transformations between pure two-party quantum states under local quantum operations and classical communication~\cite{monotones}. In this sense, $C_3$ and $C_4$ are manifestly increasing functions of entanglement monotones.  

\begin{widetext}
\begin{center}
\begin{figure}[tb]
{\includegraphics[width=1.3in, angle=-90]{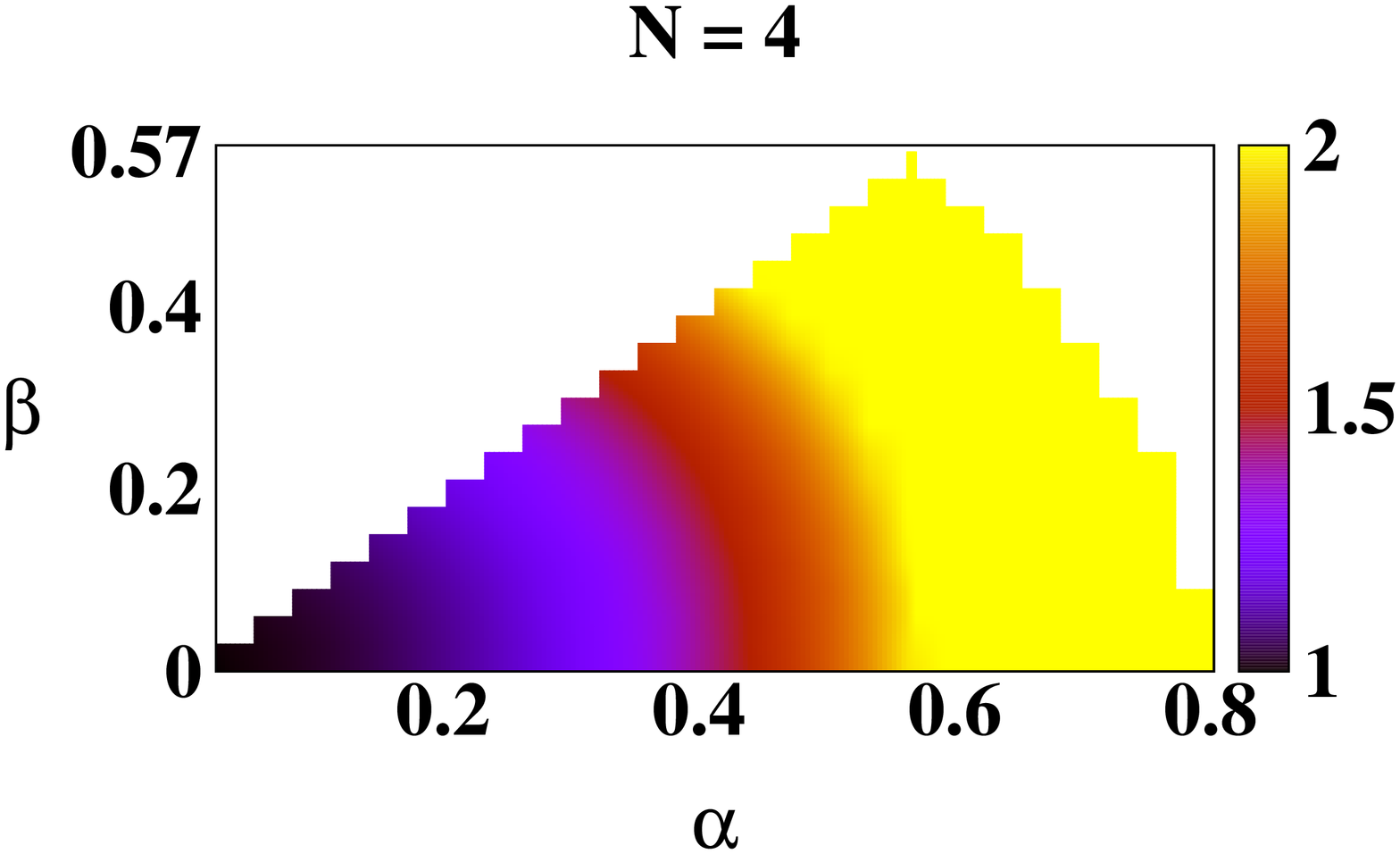}}
{\includegraphics[width=1.3in, angle=-90]{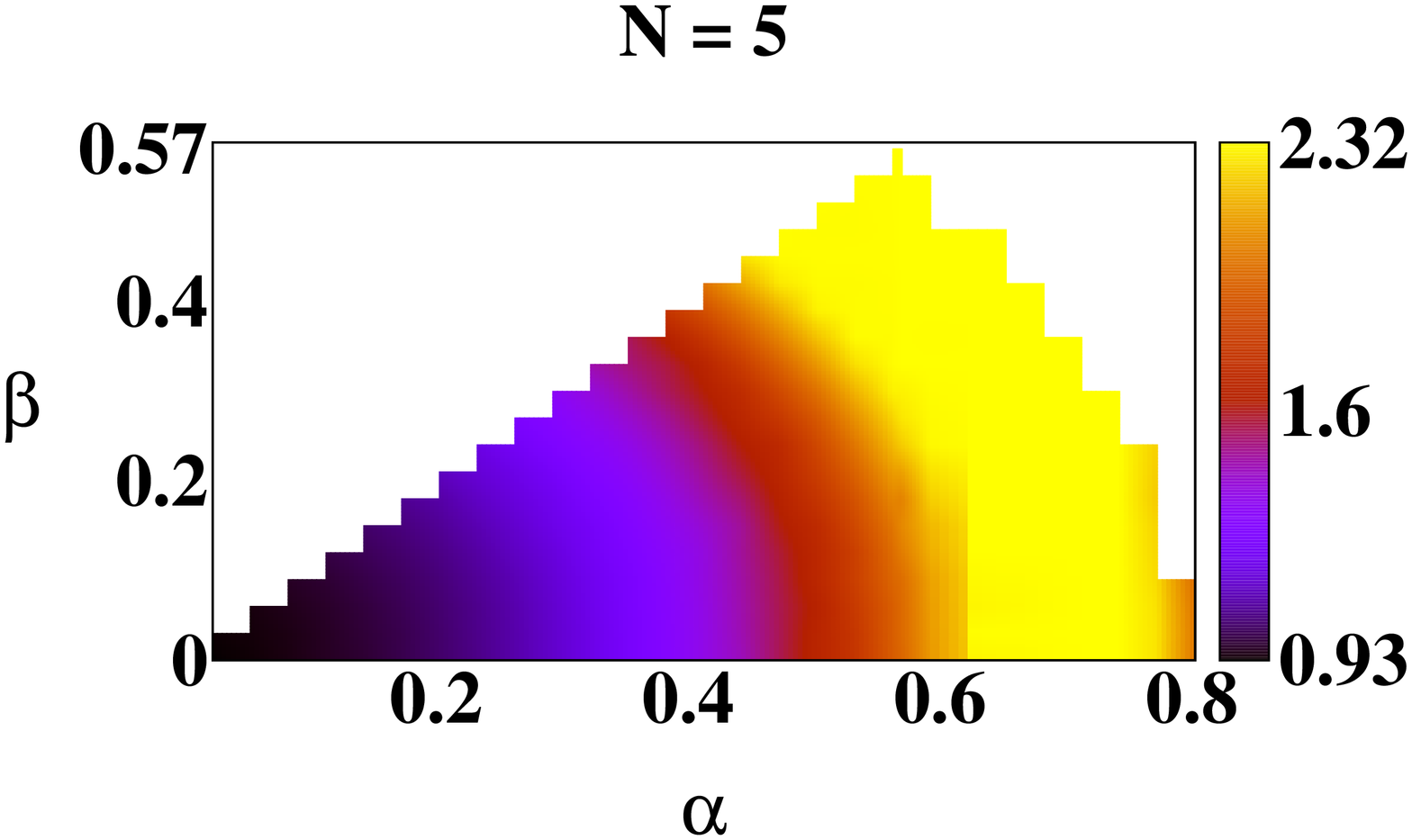}}
{\includegraphics[width=1.3in, angle=-90]{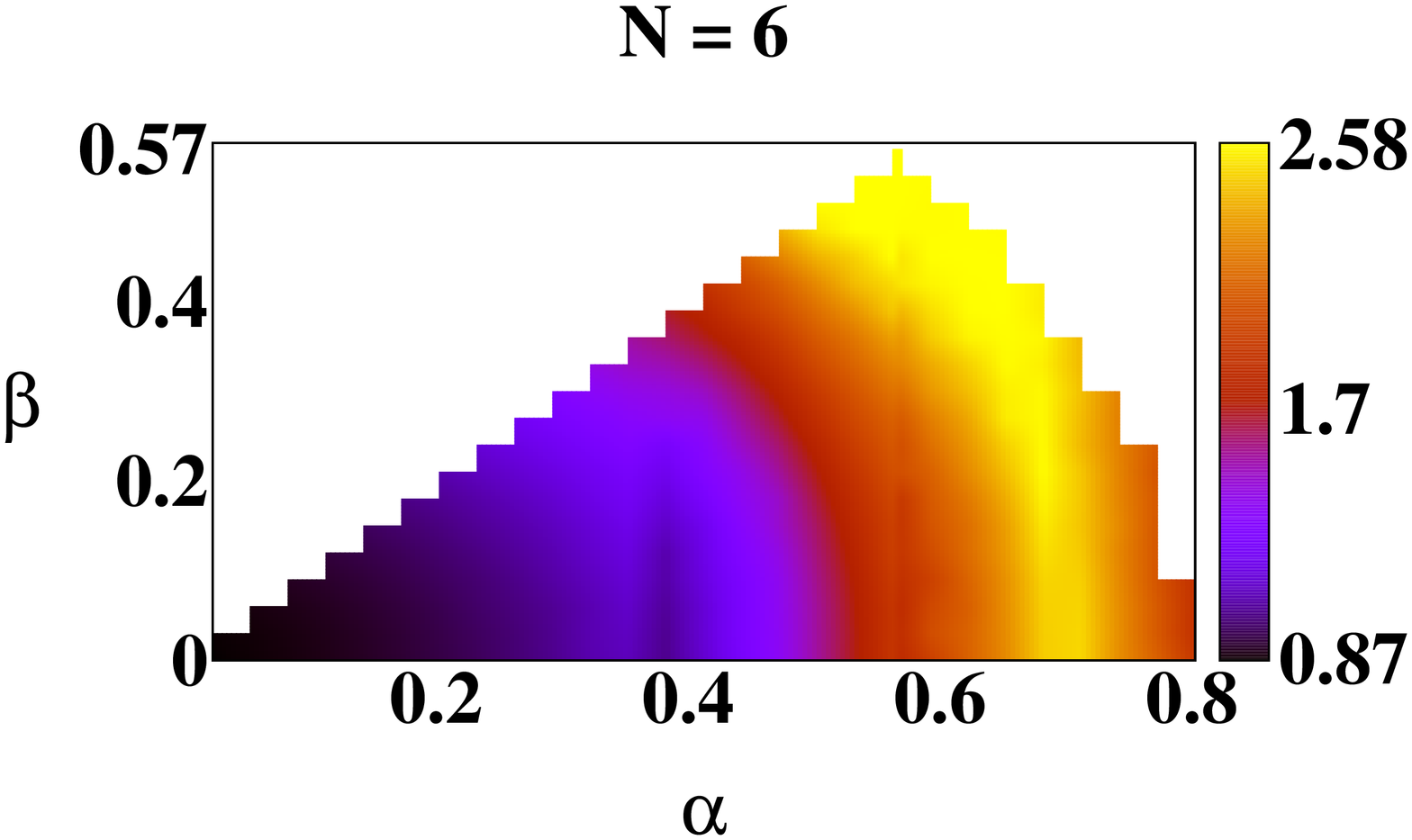}} \\
{\includegraphics[width=1.3in, angle=-90]{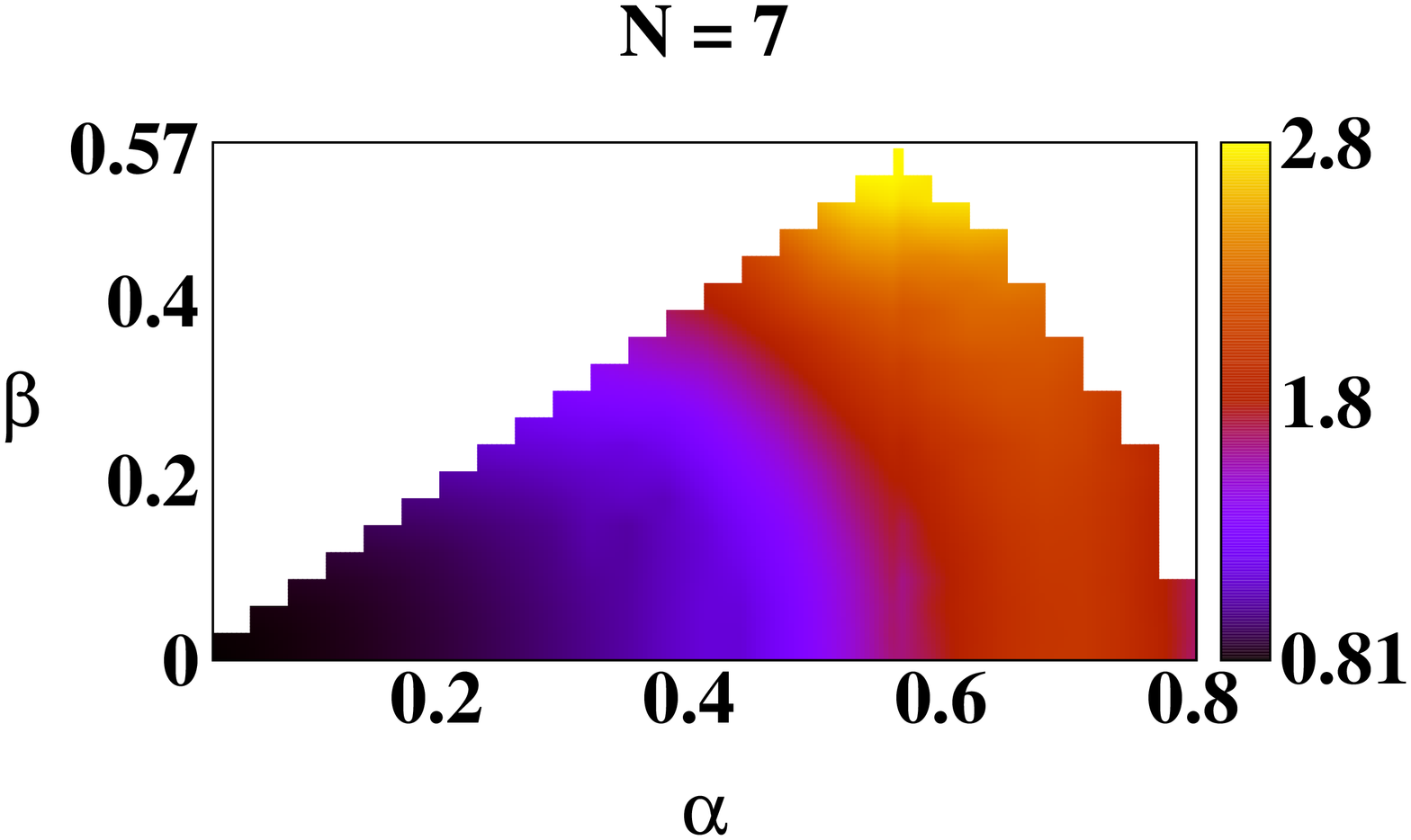}}
{\includegraphics[width=1.3in, angle=-90]{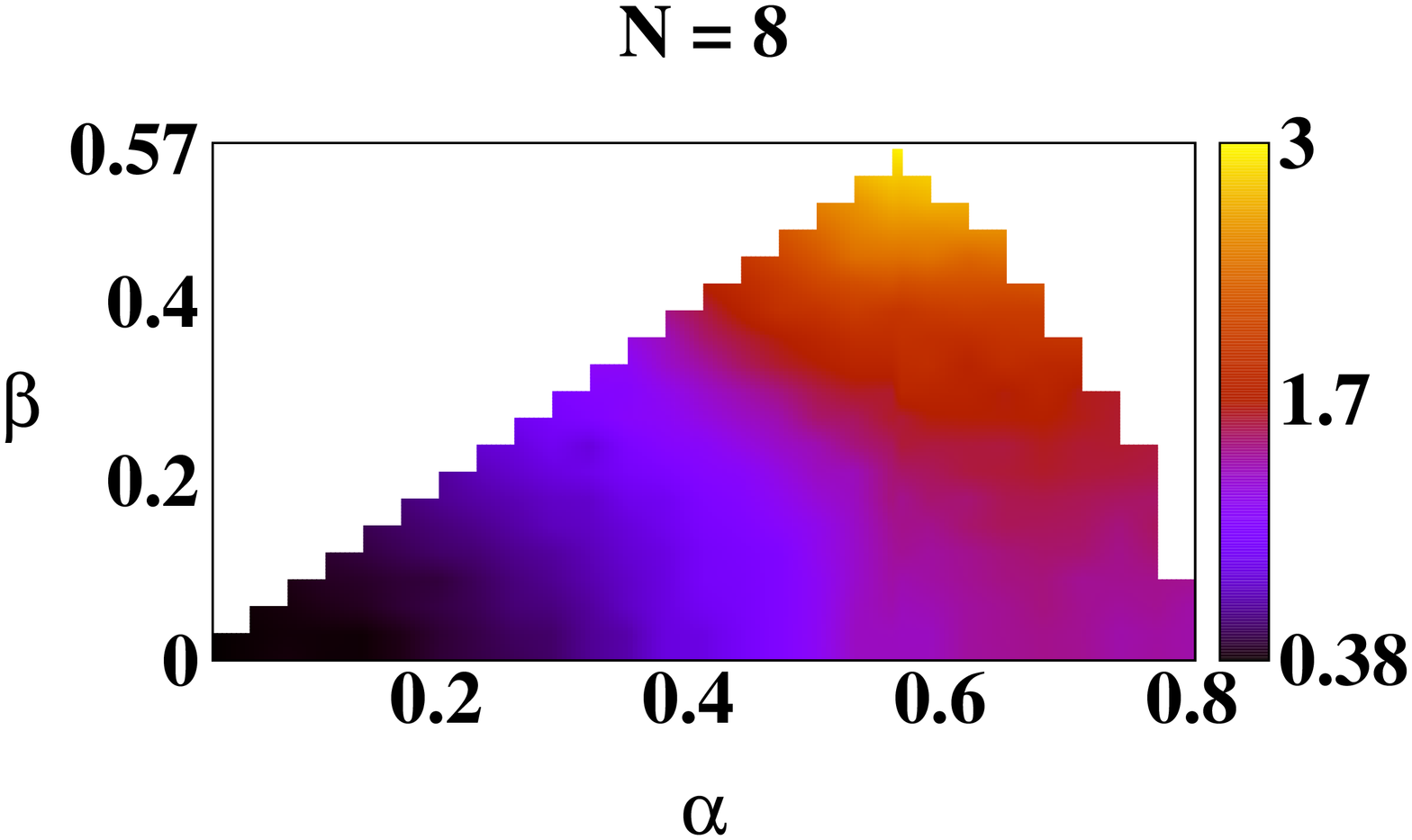}}
{\includegraphics[width=1.3in, angle=-90]{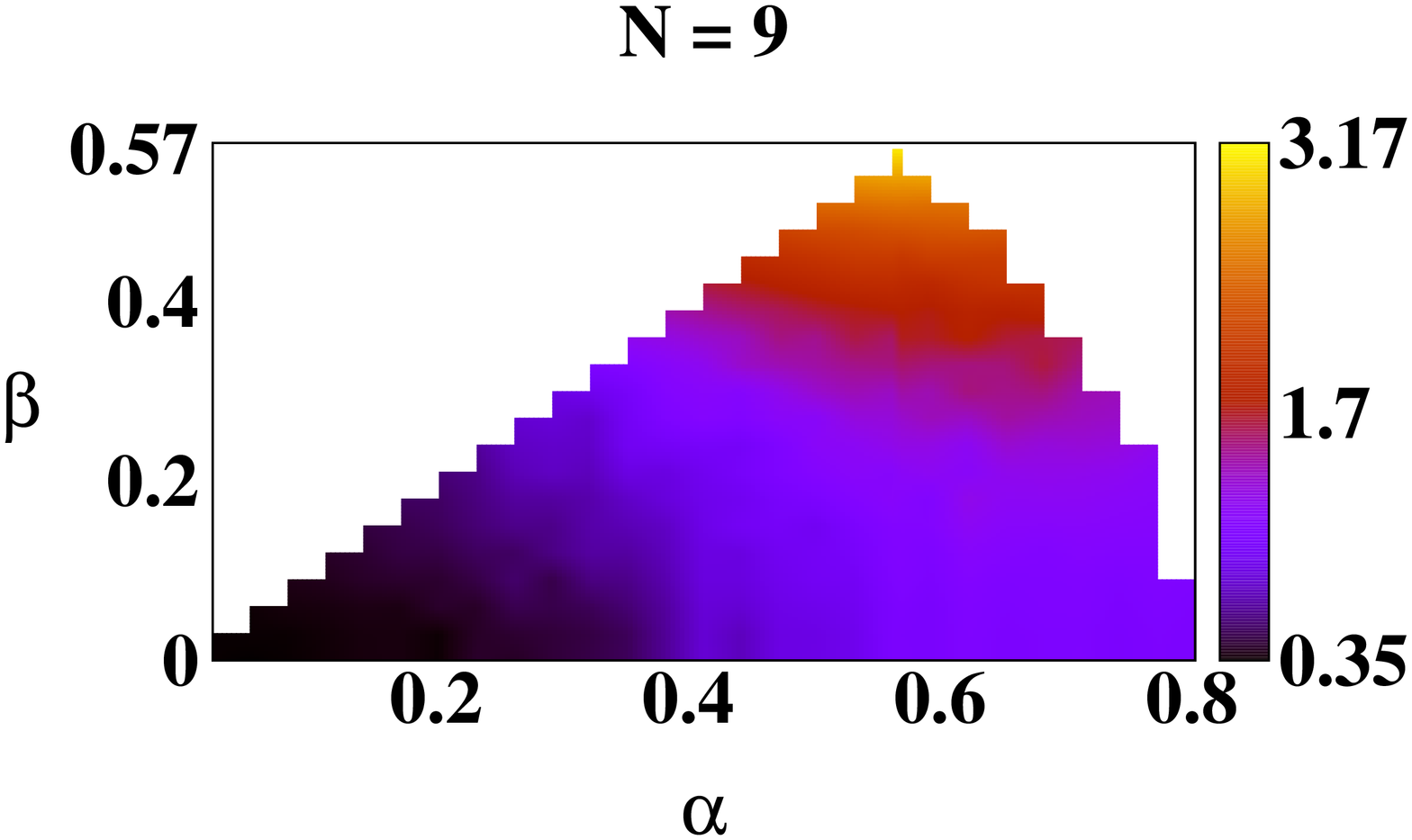}}
%
%
\caption{Conclusive quantum dense coding capacities for two qutrits. The CDC capacities $C_N$ in $\mathbb{C}^3 \otimes \mathbb{C}^3$ are plotted for $N=4,5,\ldots,9$ against the state parameters $\alpha$ and $\beta$. Clearly all entangled states are conclusively dense codable, i.e. $C_N>0$ for each $N$. The bright yellow regions represent states which are deterministically dense codable, i.e. $C_N=\log_2{N}$. All quantities are dimensionless, except the capacities, which are in bits.}
\label{qutrit}
\end{figure}
\end{center}

\begin{center}
\begin{figure}[h]
      \includegraphics[width=1.4in, angle=-90]{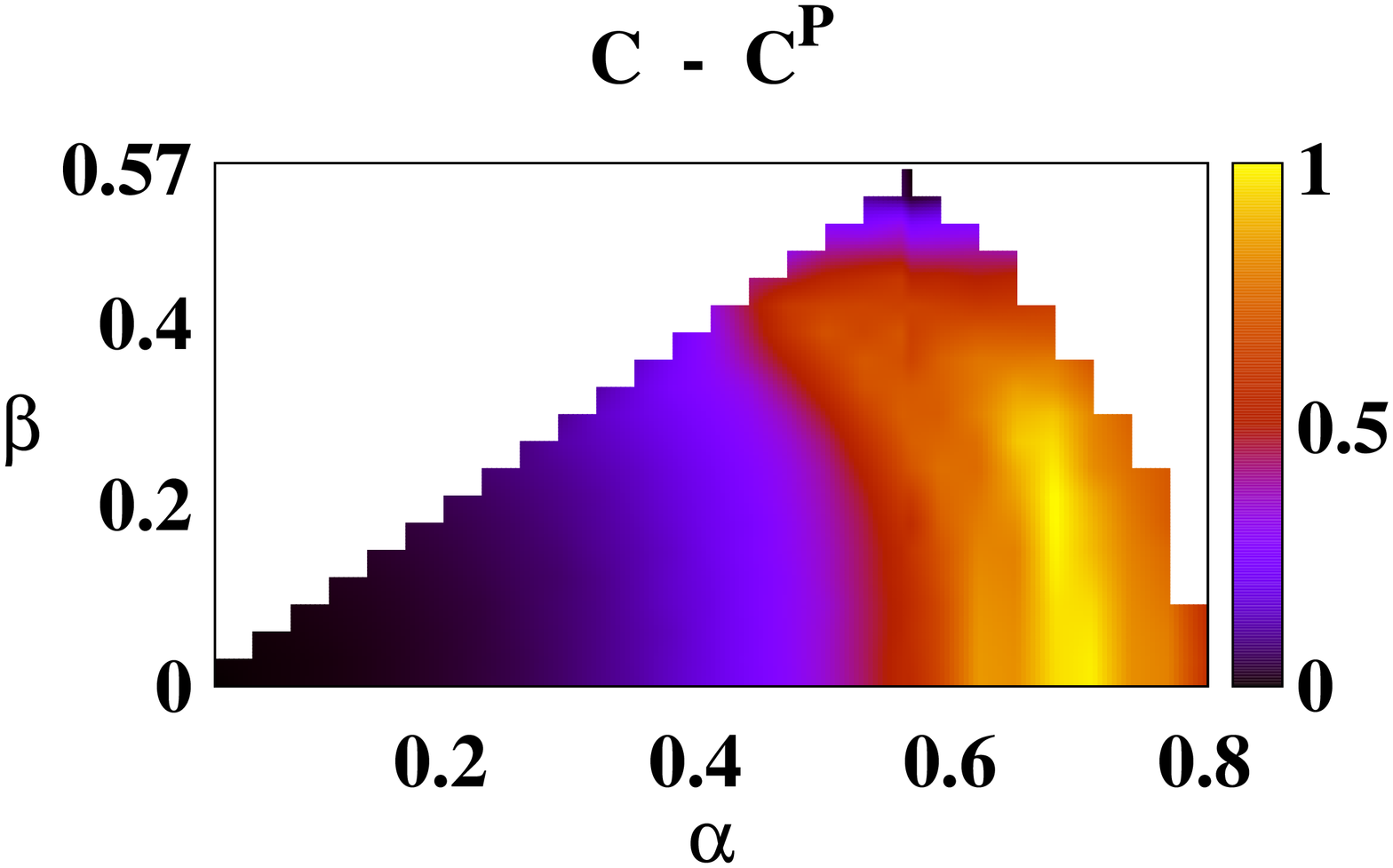}
      \includegraphics[width=1.4in, angle=-90]{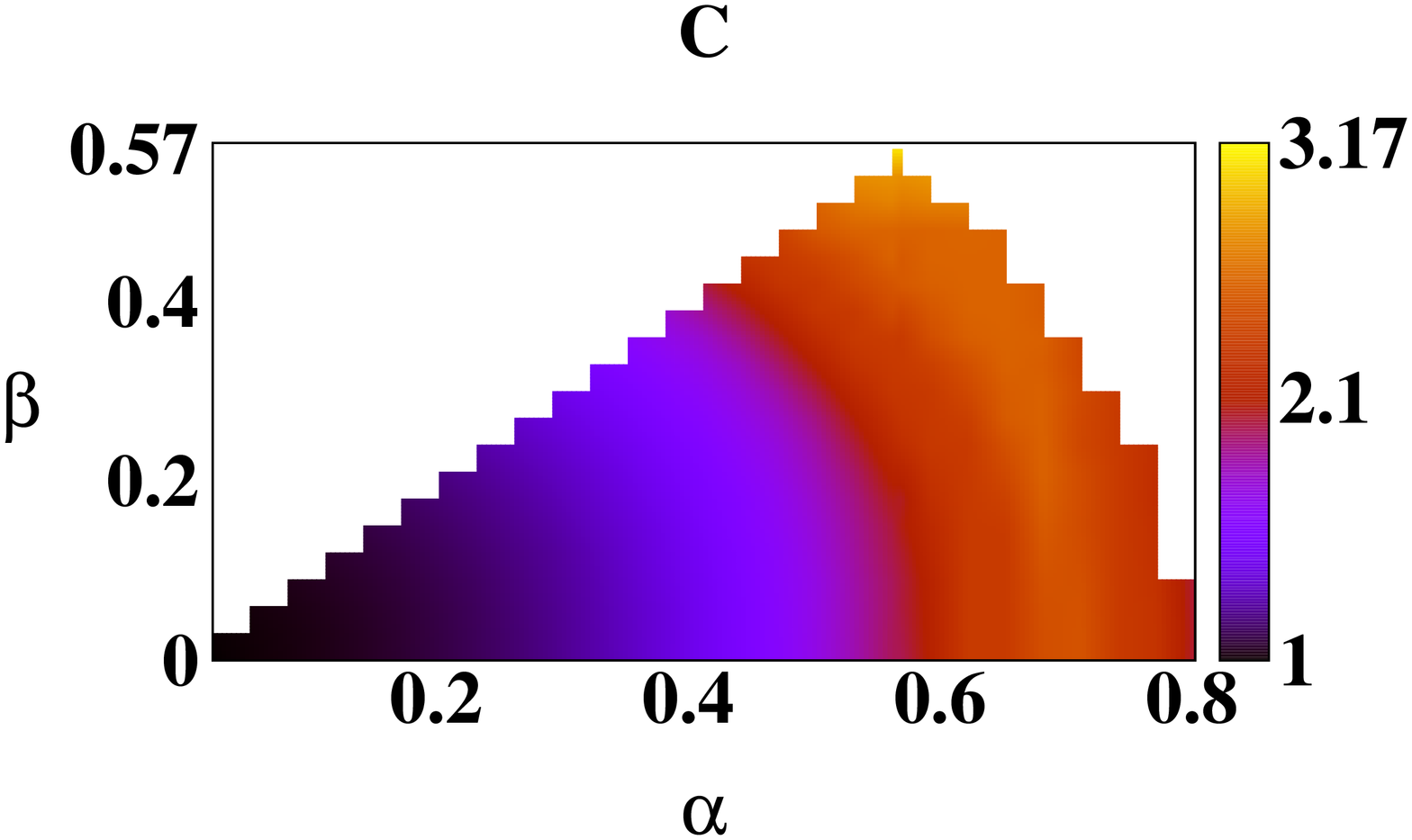} 
      \includegraphics[width=1.4in, angle=-90]{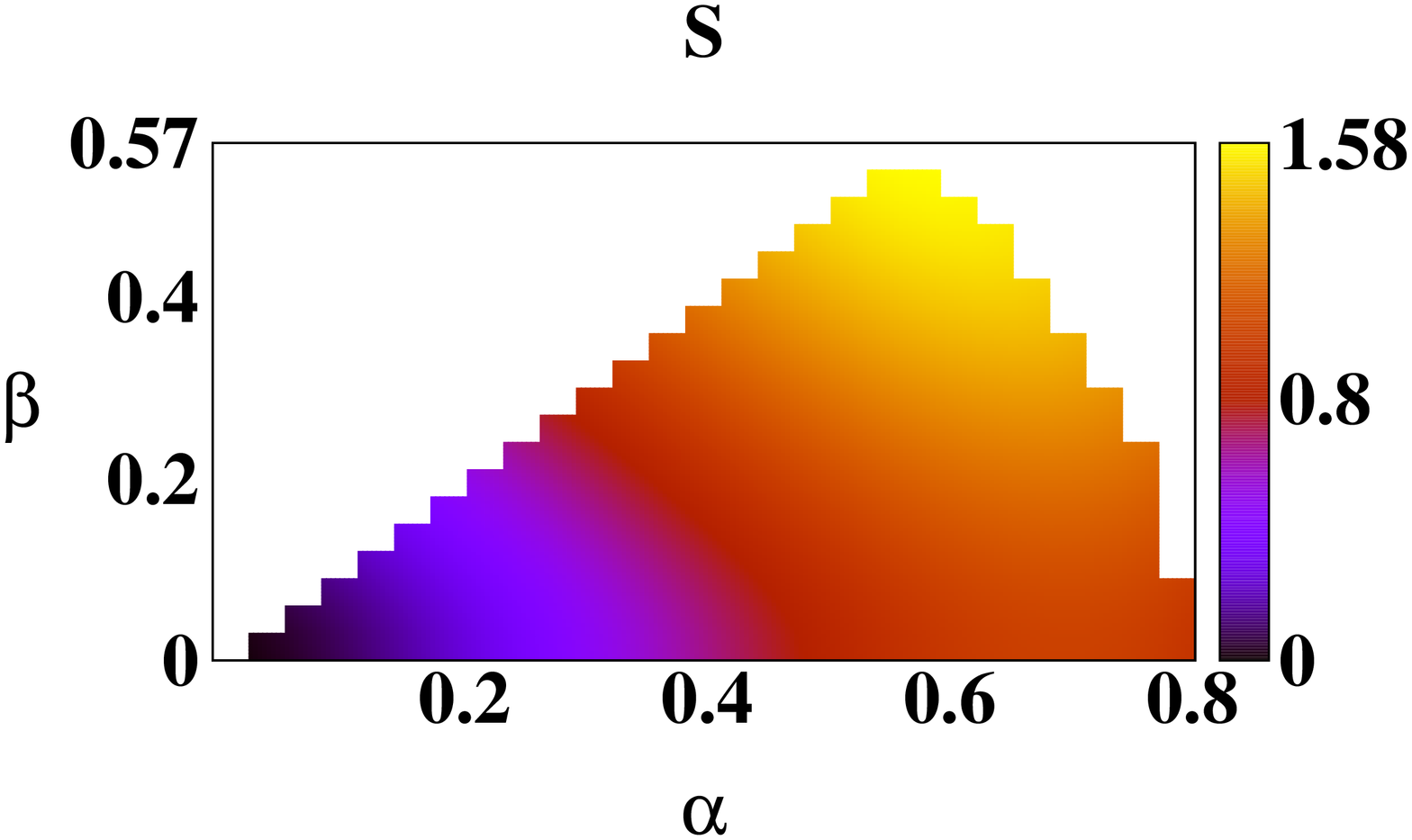}
\caption{Conclusive quantum dense coding for two qutrits. On the left panel, we plot the difference $C-C^P$ against the state parameters. Clearly, there are non-maximally entangled states for which $C-C^P>0$.
The CDC capacity, $C$, maximized over all $N>3$, and the entanglement entropy, $S$,  are plotted against $\alpha$ and $\beta$ in the middle panel and on the right respectively. The plotted applicates in the left and and middle panels are in bits, while the same on the right is in ebits. The other quantities are dimensionless.}
\label{fig_3}
\end{figure}
\end{center}  
\end{widetext}

\subsection{CDC protocol in $\mathbb{C}^3 \otimes \mathbb{C}^3$}   
\label{subsec-eijey}

Let us now move to an arbitrary pure bipartite state in in $\mathbb{C}^{3} \otimes \mathbb{C}^{3}$, given by 
$|\Psi^{(3)}\ket=\alpha|00\ket+\beta|11\ket+\sqrt{1-\alpha^2-\beta^2}|22\ket$,
where $\{|0\ket,~|1\ket,~|2\ket\}$ forms an orthonormal basis in $\mathbb{C}^3$. Without loss of generality, we choose the coefficients of $|00\ket,~|11\ket,~|22\ket$ in $|\Psi^{(3)}\ket$ as real and positive,   $\alpha \leqslant \beta$, and $\alpha^2+\beta^2 \leqslant \frac{2}{3}$. The conclusive dense coding capacities $C_N$ are plotted in Fig.~\ref{qutrit} for $N = 4, 5,\ldots, 9$ with respect to state parameters $\alpha$ and $\beta$. We see that for $N=8,9$, $C_N$ does not reach $\log_2{N}$ except for the maximally entangled state ($\alpha=\beta=\frac{1}{\sqrt{3}}$), so that DDC is not implementable in these cases, confirming the results obtained in \cite{jon, ji}. However, for $4 \leqslant N \leqslant 7$,
$C_N$ does reach $\log_2{N}$ even for non-maximally entangled states (the bright yellow regions in the panels in Fig. (\ref{qutrit})).
We also find that the area  in parameter space  for which the corresponding states render themselves for DDC decreases with  increasing $N$.

For fixed \(N\), the computation of \(C_N\) and \(C\) reveals that their optimal values are not obtained from generalized Pauli operators, but from other unitaries, showing that \(C^P_N \neq C_N\), which is in contrast to the two-qubit case.
To visualize the gap, we plot the quantity \(C-C^P\) against \(\alpha\) and \(\beta\)
in Fig. \ref{fig_3}.
%
Another observation is that $\frac{C^P}{\log_2{N}}$ does not reach unity for non-maximally entangled states, implying that generalized Pauli operators are not  enough to implement DDC in non-maximally entangled states.

In \(\mathbb{C}^3 \otimes \mathbb{C}^3\), we moreover find that the connection between entanglement and the capacities of CDC is far richer than the same for 
states in \(\mathbb{C}^2 \otimes \mathbb{C}^2\). First note that the entanglement monotones for single-copy transformations for the state $|\Psi^{(3)}\ket$ are $\alpha^2+\beta^2$ and $\alpha^2$. For asymptotic transformations, the entanglement can be quantified by the local von Neumann entropy. A look at the panels  for $N=4,~N=5,~$and $N=6$ in Fig.~\ref{qutrit} tells us that $C_4,~C_5,$ and $C_6$ broadly behave as functions of $\alpha^2+\beta^2$. On the other hand, $C_7,~C_8,$ and $C_9$ (see the bottom row of panels in Fig.~\ref{qutrit}), and $C$ (middle panel in Fig.~\ref{fig_3}) broadly behave as local von Neumann entropy, plotted in the right panel in Fig.~\ref{fig_3}. In case of the latter instances, the projected capacities are convex around $\alpha=\beta=0$ until a certain $\alpha^2+\beta^2$, after which they concave outwards.

 \begin{center}
\begin{figure}[h]
\includegraphics[width = 0.4\textwidth]{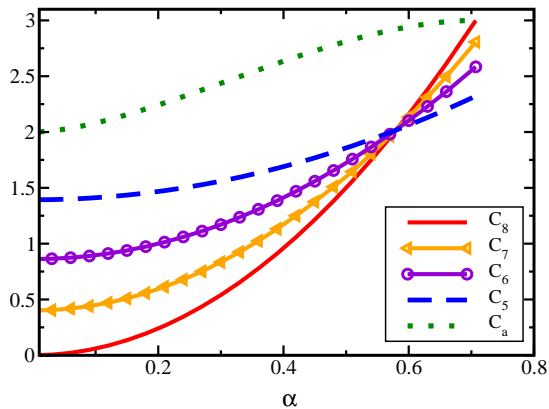} 
\caption{Multiport conclusive dense coding capacities for generalized GHZ states. The CDC capacities, $C_N$, for $N=5,6,7,8$ are plotted with respect to the state parameter $\alpha$ for the $|gGHZ\rangle$ state, with the shared state being utilized for the protocol by two senders and one receiver. Clearly, all states that are entangled in the sender to receiver bipartition, show quantum advantage. Although the curves for CDC capacities seem to be concurrent in the figure, these are actually small differences between the points of intersection of the different pairs. If we are not very close to these point of intersections, the capacity $C$ is attained by $C_5$ or $C_8$. 
The capacity $C_a$ is also plotted, which is an upper bound of the CDC capacity. All the capacities are increasing functions of entanglement in the senders : receiver bipartition as well as the GGM of the shared state. The vertical axis is in bits, while the horizontal axis is dimensionless.}
\label{fig.4}
\end{figure}
\end{center}

\section{Multiport Conclusive dense coding}
\label{multipartyPDC}


Until now, we have 
been discussing the protocol of conclusive dense coding with a single sender and a single receiver. Point to point communication is however not the reasonable one in all instances, and in this section, we consider the case of performing the CDC protocol with two senders and a single receiver, each situated at a separate location. We suppose that the three parties share a three-qubit pure state, with one qubit at each party. We study the CDC capacity, $C_N$, of such a state, with which the senders aim to send an $N$-level classical message to the receiver. Additionally, there are two noiseless qubit channels from the two senders  to the receiver. If the parties do not share an entangled state in the senders : receiver bipartition, then the senders can send at most a 4-level message, even if the senders are allowed global unitaries. We will however be interested in the scenario where only local unitaries are possible to be utilized by the senders for encoding the $N$-level classical message. Let us call the senders as Alice 1 and Alice 2, denoting them respectively as $A_1$ and $A_2$, while the receiver is still called Bob ($B$). In case the senders wish to send a classical message of $N>4$ levels, application of $N>4$ local unitaries, $U_i^{A_1A_2}=\bar{U}_i^{A_1} \otimes \bar{\bar{U}}_i^{A_2}$, with $i=1,2,\ldots,N$, is necessary, with Alice 1 applying the $\bar{U}_i^{A_1}$ and Alice 2 the $\bar{\bar{U}}_i^{A_2}$. Suppose now that the shared state is $|\Psi\ket_{A_1 A_2 B}$. We again deal with the case of equiprobable classical messages, so that after the action of the local unitaries by two Alices, the resulting ensemble consists of the states $|\Psi_i\ket_{A_1 A_2 B} = \bar{U}_i^{A_1} \otimes \bar{\bar{U}}_i^{A_2} \otimes \mathbb{I}^B |\Psi\ket_{A_1 A_2 B}$, with equal probabilities. In case $|\Psi\ket$ is a product state in the $A_1 A_2:B$ bipartition, and $N>4$, the $|\Psi_i\ket$ will be linearly dependent for any choice of the unitaries by the Alices (even global ones), and a conclusive transfer of information will not be possible, as a conclusive distinguishing of linearly dependant states is forbidden in quantum mechanics~\cite{Duan}.
 We are, therefore, interested in studying the CDC capacity, $C_N$, for shared state which is entangled in the senders : receiver bipartition, and where $5\leqslant N \leqslant 8$. Furthermore, we choose to work with genuinely tripartite pure entangled states, where the adjective ``genuinely" is used to imply that the state is entangled across all bipartitions.

 
The family of three-qubit genuinely entangled pure states is a disjoint union of the paradigmatic classes 
of GHZ (Greeenberger-Horne-Zeilinger) and W. These classes are characterized by the fact that a single copy of any state within any class can be transformed into any other state within the same class by stochastic local quantum operations and classical communication (SLOCC) between the three observers, while it cannot be transformed into any state of the other class~\cite{dur}. Important sets of states within these two classes are respectively the generalized GHZ and generalized W states. Previously it has been proved that deterministic dense coding of generalized GHZ states for the non-trivial case (transfer of a more than 4-level classical message) is not possible unless the shared state is a GHZ state. However, with some specific parameter values, the DDC capacities of three-qubit generalized W states can be non-classical~\cite{sap}. 

It is also interesting to study the behavior of CDC capacity with the entanglement of the initial shared pure state. In the case of two senders and a single receiver, there are two kinds of entanglement one can watch out for. One is the genuine multipartite entanglement present in the state and other is the entanglement in the senders : receiver bipartition of the state. As a measure of genuine multiparty entanglement, we employ the generalized geometric measure (GGM)~\cite{ggm,gm}, while in the latter case, we evaluate entanglement entropy in the senders : receiver bipartition. In this section, we first present a definition of GGM, and then study the CDC capacity of generalized GHZ and W states. Finally, we relate their behavior with GGM and entanglement entropy in the senders : receiver bipartition. 

\begin{widetext}
\begin{center}
\begin{figure}[ht]
      \includegraphics[width=1.6in, angle=-90]{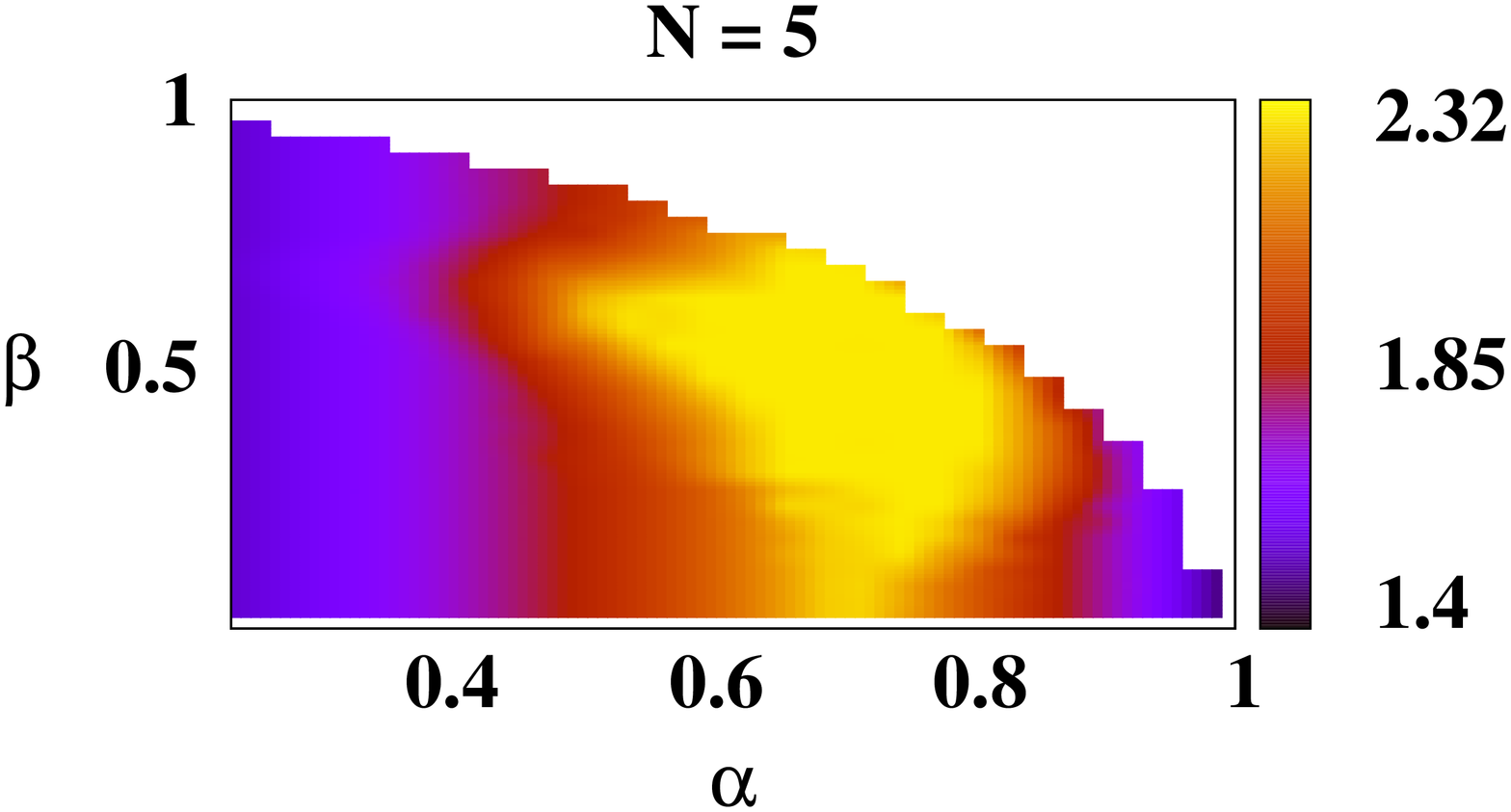}
      \includegraphics[width=1.6in, angle=-90]{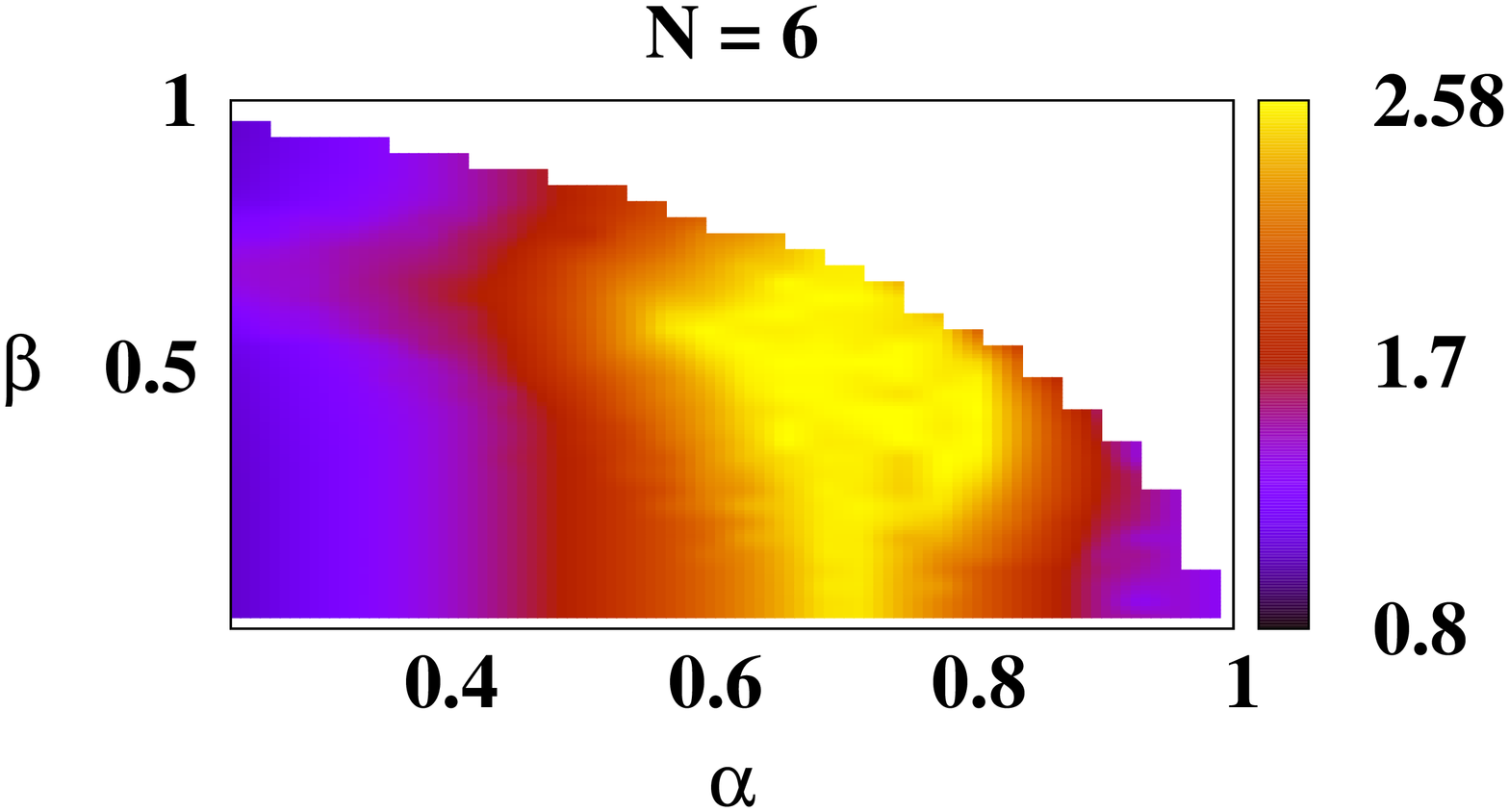} \\
      \includegraphics[width=1.6in, angle=-90]{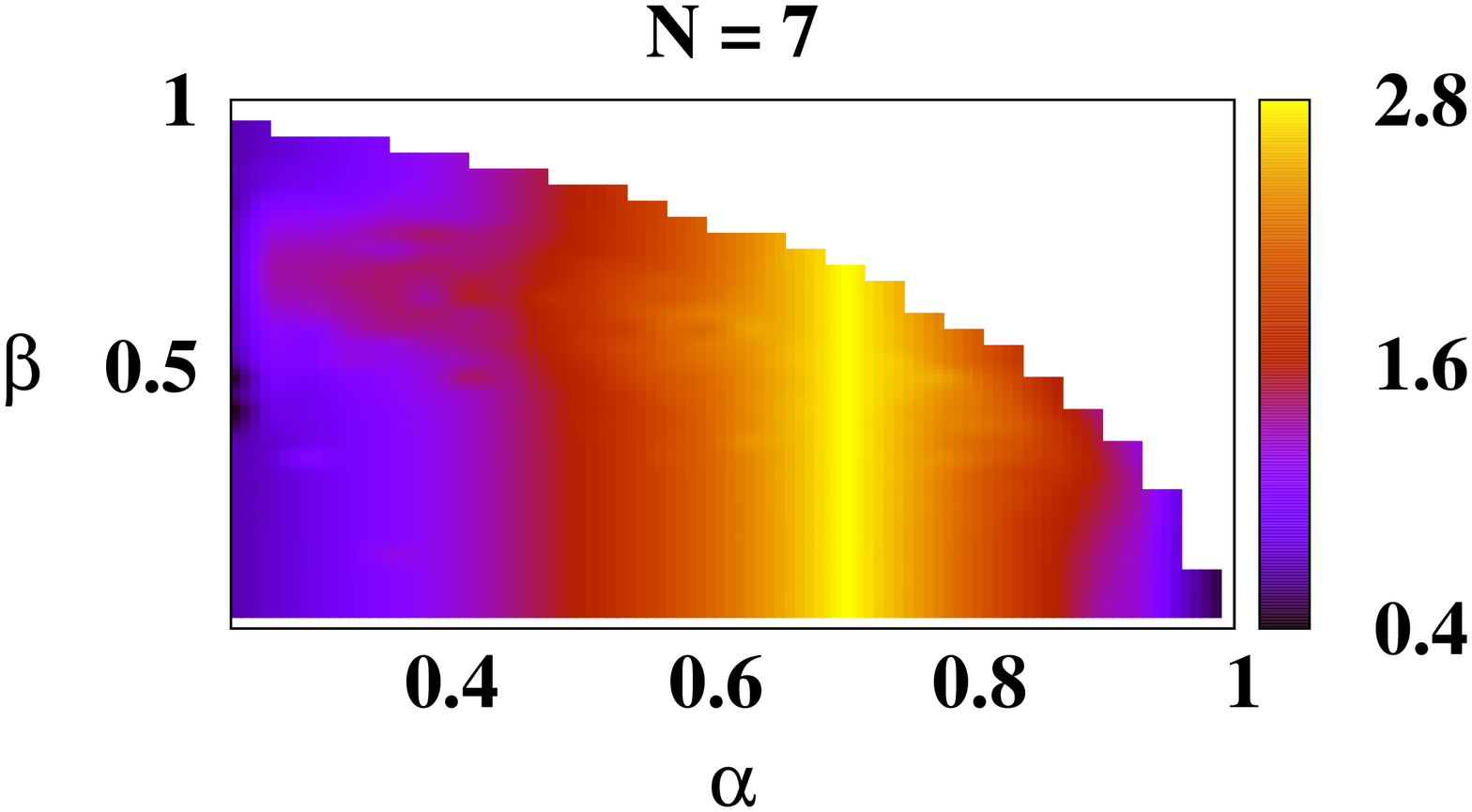}
      \includegraphics[width=1.6in, angle=-90]{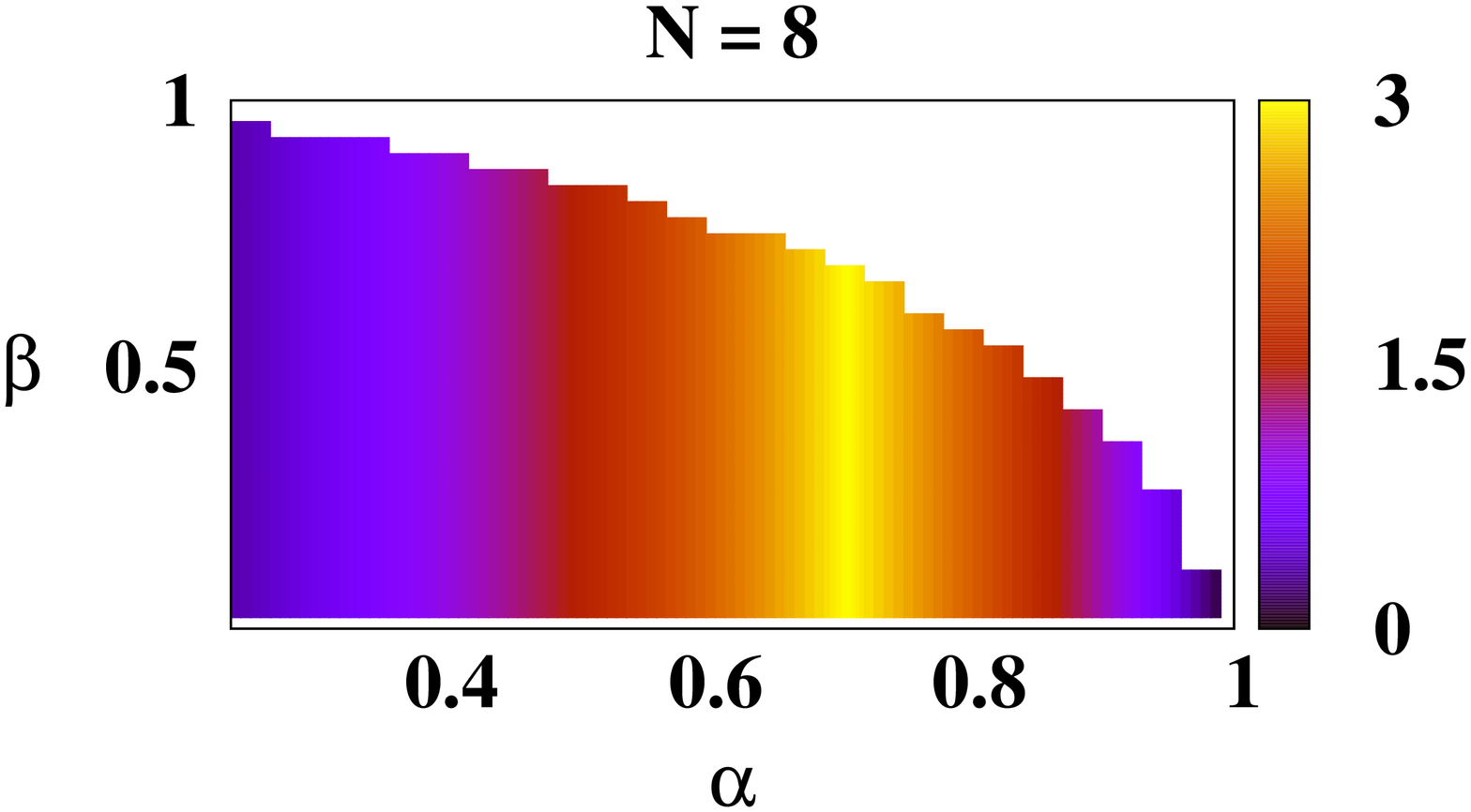}
\caption{Multiport conclusive dense coding capacities for generalized W states. The CDC capacities $C_N$  are plotted against the state parameters $\alpha$ and $\beta$ for $N=5$, 6, 7, and 8. The CDC capacity, $C_8$, is independent of $\beta$, but $C_5$, $C_6$, and $C_7$ depend on it. For each panel, the bright yellow color in the parameter space implies that DDC is executable, i.e. $C_N=\log_2{N}$, for states corresponding to such parameters. For $N=7$ and 8, there exists only a straight line, $\alpha=\frac{1}{\sqrt{2}},$ in parameter space, where DDC can be performed. But for $N=5$ and 6, there exist states other than the states on that line, for which DDC can be implemented. The capacities are in bits, while $\alpha$ and $\beta$ are dimensionless.}
\label{gw_cap}
\end{figure}
\end{center}
\end{widetext}

\begin{center}
\begin{figure}[h]
\includegraphics[width=2.3in, angle=-90]{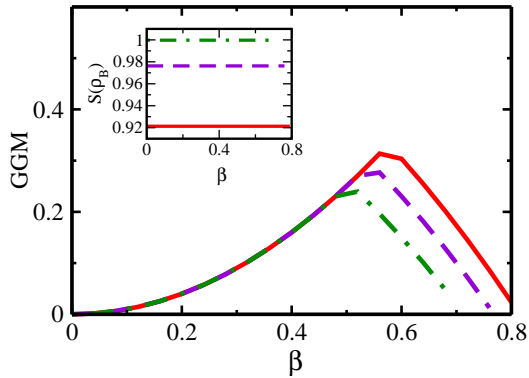}
\caption{Bipartite and genuine multipartite entanglement of generalized W states. The GGM, a genuinely multiparty entanglement measure, is plotted for $|gW\ket$ against the state parameter $\beta$ for $\alpha=0.58,0.64$, and 0.7, and they are represented in the figure by red solid, violet dashed, green dashed-dotted lines respectively. In the inset, we plot the entanglement entropy in the \(A_1A_2:B\) bipartition for the same $\alpha$ values with respect to $\beta$. Note that the entanglement entropy in the senders : receiver bipartition is independent of $\beta$. The vertical axis in the inset is in ebits. All other axes are dimensionless.}
 \label{ggmwfig}
\end{figure}
\end{center}

 \subsection{Generalized geometric measure}
 \label{ggm-defn}
  
A multipartite pure state $|\psi\ket$ is said to be genuinely multiparty entangled if it is entangled in each of the possible bipartitions of the parties. 
The GGM of $|\psi\ket$ is given by 
  \begin{equation}
  \label{Eq.11}
  \mathcal{E}(|\psi\rangle) = 1 - \Lambda_{max}^{2}(|\psi\rangle),
  \end{equation}
  where $\Lambda_{max}(|\psi\rangle)=\max_{|\phi\rangle}|\bra\phi|\psi\ket|$, the maximum being over all pure states $|\phi\ket$ which are not genuinely multiparty entangled.
 Further the quantity $\Lambda_{max}(|\psi\rangle)$ is found to have a simple form  by considering Schmidt decompositions of $|\psi\rangle$ in bipartite cuts, so that
 \begin{equation}
 \label{Eq.12}
  \mathcal{E}(|\psi\rangle) = 1 -\max \{\lambda_{\bar{A}:\bar{B}} | \bar{A}\cup \bar{B}=\{1, 2,\ldots\}, \bar{A} \cap \bar{B} =\emptyset\},
 \end{equation}
 where $\lambda_{\bar{A}:\bar{B}}$ is the largest eigenvalue of the marginal density matrix $\rho_{\bar{A}}$ or $\rho_{\bar{B}}$ of $|\psi\rangle$.\\
 

\subsection{CDC capacity with generalized GHZ state}
\label{gGHZ_capacity}
Suppose that the three parties,  Alice 1, Alice 2 and Bob share a three-qubit generalized GHZ state, 
\begin{equation}
|gGHZ\ket_{A_1A_2B}=\alpha|000\ket+\sqrt{1-\alpha^2}|111\ket.
\end{equation}
Without loss of generality, we consider $\alpha$ as real and $0 \leqslant \alpha \leqslant \frac{1}{\sqrt{2}}$. For $\alpha=\frac{1}{\sqrt{2}}$, we obtain the GHZ state. In Fig.~\ref{fig.4}, we plot the CDC capacity, $C_N$, for $N = 5, 6, 7$, and $8$ (the non-trivial cases). Clearly, quantum advantage is seen for all non-zero values of the state parameter $\alpha$. In case the senders : receiver bipartition is maximally entangled $(\alpha=\frac{1}{\sqrt{2}})$, i.e., we have the GHZ state, $C_N=\log_2{N}$, indicating that DDC is executable therein. But for the generalized GHZ state with $\alpha<\frac{1}{\sqrt{2}}$, $C_N<\log_2{N}$, and hence DDC is not possible~\cite{sap}.

Consider now the CDC capacity, $C_N^P$, obtained by using Pauli operators as encoding unitaries by Alice 1 and Alice 2. One can find the analytical form for $C_N^P$, and it is given by 
\begin{equation}
\label{Eq.13}
C_N^P=\frac{8-N+2\alpha^2(2N-8)}{N}\log_2{N},
\end{equation}
for $N=5,6,7$ and 8.

We find that the CDC capacity, $C_N$, obtained by using $arbitrary$ unitaries as encoders, coincides with that using Pauli operators and hence $C_N=C_N^P$. 
And thus, on maximization over the relevant values of $N$, we get
\begin{equation}
C=C^P.
\end{equation}
 Interestingly, all the CDC capacity curves ($C_N$), when plotted against $\alpha$, seem to intersect each other at the same point, but actually all of them intersect each other within a small range of $\alpha$ near $\alpha=0.57$. See Fig.~\ref{fig.4}.  
By using Eq.~(\ref{Eq.13}), it can be shown that there exist no $\alpha$ for which $C_N=C_{N^{'}}$ and $C_N=C_{N^{''}}$ can hold simultaneously, where no two among $N,~N,'$ and $N''$ are the same.

The  DC capacity in this case is given by $C_a=2+S(\rho_B)$, which differs from entanglement entropy, $S(\rho_B)$, in the \(A_1A_2:B\) cut, by just an additive constant.  Clearly $C_a$ is an increasing function of entanglement entropy in the senders : receiver bipartition. Since all $C_N$ are increasing functions of the state parameter $\alpha$, like the two-qubit case, 
the CDC capacities, $C_N$, are also increasing functions of entanglement entropy.
Also, as noted before in Sec. \ref{2-2 system}, $\alpha$ is itself a measure of entanglement of $|gGHZ\rangle$ in the $A_1 A_2:B$ bipartition. Moreover, for the three-party $|gGHZ\rangle$ state,  the GGM is given by
\begin{equation}
\mathcal{E}(|gGHZ\rangle) = \min\{\alpha^2,1-\alpha^2\}.  
\end{equation} 
 Therefore, for $\alpha \leqslant \frac{1}{\sqrt{2}}$, GGM increases with $\alpha$. Hence the CDC capacities, $C_N$, are also increasing functions of the GGM of the generalized GHZ state. 
 We therefore find that the GGM and entanglement entropy are rather simply related 
 to the CDC capacities. We will now show that such an uninvolved connection is 
 no longer true for general tripartite states.

An interesting feature that we have encountered till now by studying the case of parties sharing a pure entangled state in bipartite setting or a generalized GHZ state in tripartite setting is that in both the cases, the CDC capacities were obtained merely by encoding with Pauli operators. This motivates the study of generalized GHZ states for more than three parties. Consider therfore the setting where there are three senders ($A_1,~A_2,~A_3$) and a single receiver ($B$) sharing a 4-party generalized GHZ state,
$|gGHZ\ket_{A_1A_2A_3B}=\alpha|0000\ket+ \sqrt{1-\alpha^2}|1111\ket$. We obtain the CDC capacity, $C_N$, for sending an $N$-valued classical message using arbitrary unitary encoding, where $N>8$, by using numerical methods. We also obtain CDC capacity, $C^P_N$, using only Pauli operators in encoding. Interstingly, in this case too, $C_N=C^P_N$, and the analytical form is given by 
\begin{equation}
\frac{16-N+2\alpha^2(2N-16)}{N}\log_2{N}.
\label{new1}
\end{equation}
Looking at (\ref{Eq.9}), (\ref{Eq.13}), and (\ref{new1}) it seems plausible that if there are $n$ parties sharing a multiparty state $|gGHZ\ket_{A_1,A_2,\ldots,A_{n-1},B}=\alpha|00\ldots0\ket + \sqrt{1-\alpha^2}|11\ldots1\ket$, with $n-1$ of them acting as senders and the remaining as a receiver, then the CDC capacity for sending an $N$-valued classical message will be given by
\begin{equation}
\frac{2^n-N+2\alpha^2(2N-2^n)}{N}\log_2{N},
\label{new2}
\end{equation} 
 where $N>2^{n-1}$ (see Appendix).
It is easy to check that using Pauli operators for encoding, it is possible to reach the capacity in (\ref{new2}). 

\subsection{CDC with generalized W state}
\label{gW}
In a similar protocol like in the preceding subsection, suppose now that the three parties share a generalized $W$ state,
\begin{equation}
|gW\ket_{A_1A_2B} = \alpha|001\ket+\beta|010\ket+\sqrt{1-\alpha^2-\beta^2}|100\ket,
\end{equation}
where, without loss of generality, $\alpha,~\beta$ are assumed to be positive real numbers with $\alpha^2 +\beta^2 \leqslant 1$.
We find that there is quantum advantage for every state of this family, for $N =5,6,7$, and $8$. In Fig.~\ref{gw_cap}, we plot the CDC capacity $C_N$, for each $N$, with respect to the state parameters $\alpha$ and $\beta$. The states corresponding to bright yellow patches or curves, on the panels in the figure, are those for which DDC can be implemented, because they represent the situations where $C_N=\log_2{N}.$ Compare with Ref.~\cite{sap}.   
An important observation is that for $N=7$ and $8$, only the states falling on the straight line, $\alpha=\frac{1}{\sqrt{2}}$, are deterministically dense codable. Whereas, for $N=5$ and 6, along with the states on that line, there exists a neighboring patch that are also deterministically dense codable, including the W state $(\alpha=\beta=\frac{1}{\sqrt{3}})$. Another important feature is that $C_5,~C_6,$ and $C_7$ depend on both the state parameters $\alpha$ and $\beta$, but $C_8$ becomes independent of $\beta$. To understand this asymmetry between $\alpha$ and $\beta$, note that the entanglement entropy in the senders : receiver partition (i.e., in the $A_1A_2:B$ bipartition), is $H(\alpha^2) = -\alpha^2\log_2{\alpha^2} - (1-\alpha^2)\log_2{(1-\alpha^2)},$ and is therefore independent of $\beta$. The situation is different in other bipartitions, and e.g., in $A_1B:A_2$, one depends on $\beta$, but is independent of $\alpha$. Note, however, that $C_N$ for $N=5,6,7$ do depend on $\beta$ along with $\alpha$, as a consequence of which these capacities can increase while the entanglement in the $A_1A_2:B$ partition is decreasing. This is in stark contrast with generalised GHZ states, where all $C_N$ are monotonically increasing functions of  entanglement in the $A_1A_2:B$ bipartition.

Consider now the CDC capacity, $C_N^P$, of the generalized W states, where Pauli operators are used for encoding  by Alice 1 and Alice 2. It is observed that  there are states in the family of generalized W states for which $C_N > C_N^P$.
This is another marked difference observed between the CDC capacities of generalized GHZ states (for which $C_N = C_N^P$) and generalized W states.

As we have mentioned before, with respect to the protocol being considered, intuitively it seems that there are two categories of entanglement to pay attention to, viz., entanglement in the $A_1A_2:B$ partition and multiparty entanglement of the shared three-party state. We have already seen that the former, as quantified by the von Neumann entropy of the $A_1A_2$ (or the $B$) part of the shared state, is independent of $\beta$. However, except for $C_8$, the other CDC capacities do depend on $\beta$, unless we are close to $\alpha=\frac{1}{\sqrt{2}}$. See the panels in Fig.~\ref{gw_cap}. Let us now look at the behavior of GGM, a measure of genuine multisite entanglement, of the shared state. In Fig.~\ref{ggmwfig} we plot the GGM of $|gW\ket$ as a function of $\beta$ for different values of $\alpha$. We find that the GGM has an asymmetric bell-like shape, with respect to $\beta$, which becomes closer to uniform as we approach close to $\alpha=\frac{1}{\sqrt{2}}$. If we now look at the plots of $C_N$ for $N=5,6,7$ in the panels in Fig.~\ref{fig_capw}, we find that they have a broadly similar behavior. With this observation, it is tempting to comment 
that the capacities of CDC depend both on the content of genuine multipartite 
entanglement and bipartite entanglement between senders and the receiver.

\begin{widetext}
\begin{center}
\begin{figure}[h]
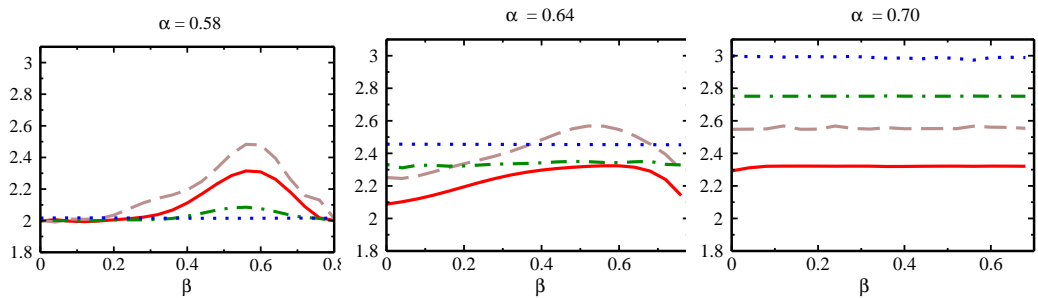


\includegraphics[width = 0.25\textwidth, angle=0]{w_pt58.eps}
\includegraphics[width = 0.25\textwidth, angle=0]{w_pt64.eps}
\includegraphics[width = 0.25\textwidth, angle=0]{w_pt7.eps}
\caption{Multiport conclusive dense coding capacities and multisite entanglement for generalized W states. The CDC capacities, $C_N$, for $N=5,6,7,$ and 8, are plotted against $\beta$ for different values of $\alpha$. $C_5$, $C_6$, $C_7$ and $C_8$ are respectively represented by red solid, grey dashed, green dashed-dotted, and blue dotted lines. The curves in these panels can be obtained  as cross-sections of the panels in Fig.~\ref{gw_cap}. We find that except $C_8$, all other CDC capacities have a behavior that is similar to that of the GGM (see Fig.~\ref{ggmwfig}), if we are away from the point of maximal entanglement ($\alpha=\frac{1}{\sqrt{2}}$) in the $A_1A_2:B$ partition. For $\alpha=\frac{1}{\sqrt{2}}$, all $C_N$ become independent of $\beta$, and behave like entanglement entropy $S(\rho_B)$ (see inset of Fig.~\ref{ggmwfig}). The vertical axes are in bits. All other quantities are dimensionless. }
\label{fig_capw}
\end{figure}
\end{center}
\end{widetext}

\section{Conclusion}
\label{conclusion}
In summary, we initiated a probabilistic version of quantum dense coding, referring to it as conclusive dense coding (CDC), when a single copy of the state is shared.
The protocol considered involves two or more than two parties, and the encoding used at the nodes of the sender/s are arbitrary unitary operators.
%
In the two-node case, we elaborated our scheme for two-qubit and two-qutrit states. 
Interestingly, we showed  that for the CDC protocol pursued by a sender and a receiver sharing a two-qutrit state, the capacity is higher if one considers arbitrary unitaries for encoding rather than  generalized Pauli operators. 
Moreover, 
we reported that although all two-qubit states except the maximally entangled one do not lend themselves for deterministic dense coding, the CDC capacity is non-classical for all two-qubit pure entangled states. We also studied the relation of entanglement of the shared state with the corresponding CDC capacity, for arbitrary  two-qubit and two-qutrit pure states. 

The multiport CDC scenario involves multiple senders and a single receiver, and the encodings by the senders were executed by arbitrary local unitaries. 
Specifically,  we found that generalized  Greenberger-Horne-Zeilinger (GHZ) as well as generalized W states provide a quantum advantage in conclusive dense coding. This is to be compared with the fact that generalized GHZ states except the GHZ state itself does not provide a quantum advantage in deterministic dense coding. We also observed that while encoding with generalized Pauli operators lead to optimal CDC capacity for generalized GHZ states, more general encoders are necessary to reach optimality for generalized W states.
Our studies revealed that 
 genuine multisite entanglement as well as bipartite entanglement in the senders : receiver bipartition of the shared states, together play roles in determining the behavior of
 the multiport CDC capacities.

 \begin{center}
 {\bf Acknowledgments}
\end{center}  
A.B. acknowledges the support of the Department of Science and Technology (DST), Government of India, through the award of an INSPIRE fellowship. We acknowledge computations performed on the cluster computing facility at the Harish-Chandra Research Institute, India.
  
 \begin{center}
 {\bf Appendix}
 \end{center}
 Here we present the proof of obtaining CDC capacity, $C^P_N,$ given in Eq. (\ref{new2}) in the case of $n$  parties sharing a generalized GHZ state, i.e., $|gGHZ\ket_{A_1A_2\ldots A_{n-1}B}=\alpha|00\ldots 0\ket + \sqrt{1-\alpha^2}|11\ldots 1\ket$, where $\alpha \leqslant \frac{1}{\sqrt{2}}$. The senders $(A_1,A_2,\ldots,A_{n-1})$ want to send an $N$-valued classical message to a single receiver $(B)$. The result for the case when $N=3$ and $n=2$ was already stated in Ref. \cite{Duan}. Senders are further restricted to use only Pauli operators for the purpose of encoding those $N$ classical messages into $N$ quantum states. Therefore, each sender can use $\mathbb{I},~\sigma_x,~\sigma_y$, and $\sigma_z$ to act on the initial shared state to form a set of at most $2^{n}$ linearly independent states. One can easily check that if $N\leqslant 2^{n-1}$, then it is always possible to transmit information deterministically, because each of the senders can use $\mathbb{I}$ and $\sigma_x$ operators to form at most of $2^{n-1}$ orthogonal states. Therefore the non-trivial cases include such $N$-valued classical messages where $2^{n-1}<N\leqslant2^n.$
   
 So, the senders' task is to construct $N$ quantum states using application of only Pauli operators on the  initial shared state. First of all, they can construct a set $2^{n-1}$ states which are mutually orthogonal by using the unitary operators from the set $\mathcal{A}= \big\{(\mathbb{I} \otimes \mathbb{I} \otimes \ldots \otimes \mathbb{I} \otimes \mathbb{I}),~(\sigma_x \otimes \mathbb{I} \otimes \ldots \otimes \mathbb{I} \otimes \mathbb{I}),~(\mathbb{I} \otimes \sigma_x \otimes \ldots \otimes \mathbb{I} \otimes \mathbb{I}),\ldots,~(\sigma_x \otimes \sigma_x \otimes \ldots \otimes \sigma_x \otimes \mathbb{I})\big\}_{A_1A_2\ldots A_{n-1}B}$, which have $\mathbb{I}$ at $B$, and either $\sigma_x$ or $\mathbb{I}$ at the senders. And the remaining $N-2^{n-1}$ states can be obtained, without loss of generality, by using any $N-2^{n-1}$ unitaries from the set $\mathcal{A}'=\big\{(\mathbb{I} \otimes \mathbb{I} \otimes \ldots \otimes \sigma_z \otimes \mathbb{I}),~(\sigma_x \otimes \mathbb{I} \otimes \ldots \otimes \sigma_z \otimes \mathbb{I}),~(\mathbb{I} \otimes \sigma_x \otimes \ldots \otimes \sigma_z \otimes \mathbb{I}),\ldots,~(\sigma_x \otimes \sigma_x \otimes \ldots \otimes \sigma_x\sigma_z \otimes \mathbb{I})\big\}_{A_1A_2\ldots A_{n-1}B}$, which has $2^{n-1}$ elements and are formed by replacing the $\sigma_x$ or $\mathbb{I}$ at $A_{n-1}$ of the elements of $\mathcal{A}$ by a $\sigma_x \sigma_z$ or $\sigma_z$ respectively.  
 
At this point, the senders send the ensemble of $N$ states to the receiver, who can distinguish among states with efficiencies $\gamma_1,\gamma_2,\ldots,\gamma_N$, subject to constraint (\ref{cond1}). Without loss of generality, the constraint matrix in this case can be divided  in blocks of $2\otimes 2$ matrices of the form $\tilde{M}_p 
=  \begin{pmatrix}
1-\gamma_{2p-1} & 1-2\alpha^2 \\
1-2\alpha^2 & 1-\gamma_{2p}
\end{pmatrix},
\quad
$ and $1\otimes 1$ scalars, $\tilde{S_q}=1-\gamma_q$, where $1\leqslant p\leqslant N-2^{n-1}$ and $2N-2^{n}+1\leqslant q\leqslant N$. 
  
Now, non-negativity of $\tilde{M}_p$ and $\tilde{S}_q$ is equivalent to the constraint (\ref{cond1}). These provide inequalities
\begin{equation}\tag{A1}
{\gamma_q \leqslant 1}
\label{A1}
\end{equation} 
and
\begin{equation}\tag{A2}
\begin{aligned}
\gamma_{2p-1}+\gamma_{2p}\leqslant &2 - 2\Big\{\Big({\frac{(1-\gamma_{2p-1})+(1-\gamma_{2p})}{2}}\Big)^2  \\
& - (1-\gamma_{2p-1})(1-\gamma_{2p}) +  (1-2\alpha^2)^2\Big\}^\frac{1}{2}.
\end{aligned}
\label{A2}
\end{equation}

The task is to obtain the CDC capacity, $C^P_N = \max_{\{\gamma_i\}}\Sigma_{i=1}^N \frac{\gamma_i}{N} \log_2{N}$.
From inequalities ($\ref{A1}$), one gets $\max\gamma_q=1$, where $2N-2^{n}+1\leqslant q\leqslant N$.
And from inequalities ($\ref{A2}$), one gets $\gamma_{2p-1}=\gamma_{2p}=\gamma$ (say), and $\max\gamma=2\alpha^2$,  where $1\leqslant p\leqslant N-2^{n-1}$. Thus we get
\begin{equation}\tag{A3}
C^P_N=\frac{2^n-N+2\alpha^2(2N-2^n)}{N}\log_2{N}.
\end{equation}


\begin{thebibliography}{99}
\bibitem{eijey}  The ``classical limit" is explicitly defined later in the paper.
\bibitem{bell-eita} These states are known as the Bell states, being given by
\(|\psi^{\pm}\rangle=\frac{1}{\sqrt{2}} (|01\rangle \pm |10\rangle)\),
\(|\phi^{\pm}\rangle=\frac{1}{\sqrt{2}} (|00\rangle \pm |11\rangle)\).
\end{thebibliography}
\end{document}